\documentclass[acmsmall,screen,nonacm]{acmart}

\AtBeginDocument{%
  }

\setcopyright{acmlicensed}
\copyrightyear{2018}
\acmYear{2018}
\acmDOI{XXXXXXX.XXXXXXX}
\acmConference[XX]{}{XX}{XX}
\acmJournal{PACMPL}
\acmVolume{1}
\acmNumber{CONF} 
\acmArticle{1}
\acmYear{2018}
\acmMonth{1}
\acmISBN{978-1-4503-XXXX-X/2018/06}

\author{Yifan Zhang}
\orcid{0009-0005-2061-0273}             
\affiliation{
  \department{Key Laboratory of High Confidence Software Technologies (Peking University), Ministry of Education; School of Computer Science}              
  \institution{Peking University}            
  \city{Beijing}
  \country{China}                    
}
\email{yfzhang23@stu.pku.edu.cn}          

\author{Xin Zhang}
\authornote{Corresponding author.}          
\orcid{0000-0002-1515-7145}             
\affiliation{
  \department{Key Laboratory of High Confidence Software Technologies (Peking University), Ministry of Education; School of Computer Science}              
  \institution{Peking University}            
  \city{Beijing}
  \country{China}                    
}
\email{xin@pku.edu.cn}          

\usepackage{adjustbox}
\usepackage{algorithm}
\usepackage[noend]{algpseudocode}
\usepackage{bm}
\usepackage{caption}
\usepackage{colortbl}
\usepackage{enumitem}
\usepackage{listings}
\usepackage{makecell}
\usepackage{multicol}
\usepackage{multirow}
\usepackage{rotating}
\usepackage{soul}
\usepackage{subcaption}
\usepackage{tikz}
\usepackage{xcolor}
\usepackage{xspace}
\usepackage{longtable}
\usepackage{tcolorbox}
\usepackage{supertabular}
\usepackage{afterpage}
\usepackage[normalem]{ulem}
\usetikzlibrary{positioning}
\definecolor{keyword}{RGB}{193,8,70}
\definecolor{truealarm}{RGB}{166,255,166}
\newcommand{\lineref}[1]{\hyperref[#1]{Line~\ref*{#1}}}

\begin{document}

\title{Beyond Imprecise Distance Metrics: Trace-Guided Directed Greybox Fuzzing via LLM-Predicted Call Stacks}

\begin{abstract}
Directed greybox fuzzing (DGF) aims to efficiently trigger bugs at specific target locations by prioritizing seeds whose execution paths are more likely to reach the targets. However, existing DGF approaches suffer from imprecise potential estimation due to their reliance on static-analysis-based distance metrics. The over-approximation inherent in static analysis causes many seeds with execution paths irrelevant to vulnerability triggering to be mistakenly prioritized, significantly reducing fuzzing efficiency. To address this issue, we propose trace-guided directed greybox fuzzing (TDGF). TDGF replaces static-analysis-based distance metrics with vulnerability-oriented execution information (referred to as guidance traces) to steer directed fuzzing: seeds whose execution paths overlap more with the guidance traces are scheduled earlier for mutation. We empirically study two representative types of guidance traces: the control-flow trace and the call-stack trace of vulnerability-triggering executions. We find that the fine-grained control-flow traces offer nearly the same guidance capability as the coarse-grained call-stack traces, while call-stack traces are also easier for large language models (LLMs) to predict. Based on this insight, we further propose a framework that leverages LLMs to predict the call stack at vulnerability-triggering time and uses it to guide DGF. We implement our approach and evaluate it against several state-of-the-art fuzzers with experiments totaling 58.4 CPU-years. On a suite of real-world programs, our approach triggers vulnerabilities 2.13$\times$ to 3.14$\times$ faster than the baselines. Moreover, through directed patch testing on the latest program versions used in our controlled experiments, our approach discovers 10 new vulnerabilities and 2 incomplete fixes, with 10 assigned CVE IDs.
\end{abstract}

\settopmatter{printfolios=true}
\maketitle
\section{Introduction}

Directed greybox fuzzing (DGF) \cite{AFLGo} is currently one of the most effective tools for triggering bugs at specific locations. For each seed, DGF collects its execution path and uses specific strategies to calculate the potential of triggering target vulnerabilities through mutation, prioritizing the mutation of seeds with higher potential. As a result, DGF can trigger target vulnerabilities more quickly compared to conventional coverage-guided greybox fuzzing (CGF) \cite{AFL}. Research related to DGF \cite{AFLGo, Hawkeye, CAFL, Windranger, DAFL, SelectFuzz, SDFuzz, Beacon} has been widely proposed and is also extensively applied in areas such as vulnerability reproduction, patch testing,  and validation of static analysis alarms.

However, existing DGF approaches are still imprecise in calculating mutation potential toward the target. Existing DGF approaches calculate potential based on complex distance metrics, such as distances on the control flow graph (CFG) and value flow graph (VFG). The computation of these distances is derived from over-approximations in static analysis, which inevitably leads to \textbf{many flows that do not exist during actual execution that triggers target vulnerabilities}. 

To address this issue, we propose replacing imprecise static-analysis-based distance metrics with vulnerability-oriented execution information to guide directed fuzzing. We call this framework trace-guided directed greybox fuzzing (TDGF). We refer to the execution information as a guidance trace, and we operationalize it as a seed-prioritization score: seeds whose executions overlap more with vulnerability-triggering behavior are scheduled earlier for mutation. Useful guidance traces should satisfy two requirements: (1) they provide reliable prioritization during fuzzing, and (2) they can be obtained or predicted with high accuracy. We conduct an empirical study of two representative forms of guidance traces: (1) vulnerability-reaching control flow and (2) the call stack at the vulnerability-triggering point. Our results show that, given ground-truth guidance traces, both substantially outperform prior static-analysis-based distance metrics. However, fine-grained control-flow traces do not yield the expected advantage, because small mutations can induce large and irregular changes in basic-block coverage, making control-flow traces \textbf{noisy} and \textbf{less reliable} for prioritization. In contrast, call-stack traces are \textbf{higher-level} and more \textbf{stable} under mutation-induced changes, providing a more reliable basis for ranking seeds. Moreover, motivated by the strong semantic and code-aware capabilities of large language models (LLMs), we investigate using LLMs to predict such guidance traces. The results show that call-stack prediction achieves an F1-score of over \textbf{0.70}, whereas control-flow prediction attains only about \textbf{0.20}, further supporting the practical advantage of call stacks.

Based on the TDGF framework, we present \textsc{Staczzer}, a framework that leverages LLMs to predict call stacks for guiding DGF. \textsc{Staczzer} first uses static analysis to construct the call graph (CG) and, based on the CG, identifies all methods that may reach the potential vulnerability location through a sequence of calls. If the source code of these methods is short, \textsc{Staczzer} submits it directly to the LLM in a single query to predict the call stack; otherwise, it uses code agents to query the codebase directly. This design maintains high call-stack prediction accuracy while significantly reducing the time overhead of code agents on small programs, and it also improves scalability when handling large codebases. Experimental results show that LLM-predicted call stacks achieve high accuracy, and the resulting acceleration of DGF closely matches the setting where the ground-truth call stacks are known. As LLM capabilities continue to advance, DGF is expected to benefit from increasingly accurate call-stack predictions on more complex programs.

We have implemented \textsc{Staczzer} on top of the greybox fuzzing framework \textsc{AFL} \cite{AFL}, and compared it against the following fuzzers: (1) \textsc{AFLGo} \cite{AFLGo}, the first DGF tool, which employs complex CFG-based distance metrics; (2) \textsc{WindRanger} \cite{Windranger}, an optimized DGF tool that builds upon \textsc{AFLGo} by computing only deviation basic blocks (DBBs); (3) \textsc{DAFL} \cite{DAFL}, the state-of-the-art (SOTA) DGF tool that leverages complex VFG-based distance metrics; (4) \textsc{G$^2$Fuzz} \cite{non-textual}, the SOTA CGF tool that leverages LLM-generated inputs; (5) \textsc{AFL} \cite{AFL}, the base framework for all fuzzers used in our evaluation. We use the same set of 41 real-world known vulnerabilities and the same evaluation metrics as in the experiments of the SOTA DGF tool \textsc{DAFL}. The total experimental time is approximately 58.4 CPU-years. The results show that \textsc{Staczzer} triggers vulnerabilities on average $2.13\times$ to $3.14\times$ faster than these baselines. Furthermore, by performing directed testing on existing patch locations in the latest versions of controlled experiment benchmarks, \textsc{Staczzer} successfully discovered 10 new vulnerabilities and 2 incomplete fixes in high-impact software GNU linker and its related Binutils components, with 10 assigned CVE IDs.

\textit{Contributions.} This paper makes the following contributions:

\begin{enumerate}
[noitemsep,topsep=0pt,parsep=0pt,partopsep=0pt]
    \item We propose trace-guided directed greybox fuzzing to address the imprecision of conventional DGF static-analysis-based distance metrics.
    \item We conduct an empirical study of two representative types of guidance traces. Counterintuitively, the results show that fine-grained vulnerability-triggering control-flow traces are not a practical choice, whereas call-stack traces are more effective.
    \item We present \textsc{Staczzer}, a framework that integrates LLM-based prediction of vulnerability-triggering call stacks to guide DGF.
    \item We show the effectiveness of \textsc{Staczzer} on a suite of real-world programs. \textsc{Staczzer} demonstrates significantly stronger vulnerability triggering performance compared to the baselines.
\end{enumerate} 
\section{Motivating Example}\label{section-2}

In this section, using CVE-2016-4489 \cite{CVE-2016-4489} in \texttt{cxxfilt} as a case study, we show that existing DGF methods use imprecise distance metrics to estimate a seed’s mutation potential toward the target, and we then show how our approach addresses this limitation.

\begin{figure}[t!]
\begin{multicols}{2}
\begin{lstlisting}[language=C, escapechar=|, morekeywords={bool}]   
char** output; |\label{line:output}|

void gnu_special(const char** mangled){
  if('0' <= **mangled && **mangled <= '9'){ |\label{line:gnu_special_if}|
    int n = 0; |\label{line:gnu_special_init}|
    while('0' <= **mangled && **mangled <= '9'){ |\label{line:gnu_special_while}|
      n *= 10; |\label{line:gnu_special_mul}|
      n += **mangled - '0'; |\label{line:gnu_special_add}|
      (*mangled)++; |\label{line:gnu_special_add1}|
    }
    memcpy(*output, *mangled, n); // CVE-2016-4489 |\label{line:gnu_special_cve}|
  }
  ...
}

void internal_cplus_demangle(const char** mangled, bool flag){
  if(flag){ |\label{line:internal_cplus_demangle_flag}|
    if(**mangled != '\0'){ |\label{line:internal_cplus_demangle_if1}|
      memcpy(*output, "cplus_marker", 12); |\label{line:internal_cplus_demangle_memcpy}|
      (*output) += 12; |\label{line:internal_cplus_demangle_add1}|
      gnu_special(mangled); |\label{line:internal_cplus_demangle_call1}|
    }
  }else{
    if(strlen(*mangled) >= 9 && strncmp(*mangled, "_GLOBAL_", 8) == 0 && !('0' <= (*mangled)[8] && (*mangled)[8] <= '9')){ |\label{line:internal_cplus_demangle_if2}|
      (*mangled) += 8; |\label{line:internal_cplus_demangle_add2}|
      gnu_special(mangled); |\label{line:internal_cplus_demangle_call2}|
    }
  }
  ...
}

int main(){
  const char** mangled = input1(); |\label{line:main_input1}|
  bool flag = input2(); |\label{line:main_input2}|
  internal_cplus_demangle(mangled, flag); |\label{line:main_call}|
  ...
}
\end{lstlisting}
\end{multicols}
\vspace{-0.25in}
\caption{Simplified code fragment from \texttt{cxxfilt} containing CVE-2016-4489.}
\label{code-cxxfilt}
\end{figure}

\subsection{A Seed Prioritization Problem in Directed Greybox Fuzzing}

\autoref{code-cxxfilt} shows simplified \texttt{cxxfilt} code for CVE-2016-4489; we modified it for brevity while preserving vulnerability-relevant semantics. \texttt{cxxfilt} demangles C++ symbols by reversing compiler name mangling into human-readable form.
From \autoref{line:main_input1}, the code reads the mangled input via \texttt{input1} into \texttt{mangled}, reads a boolean \texttt{flag} via \texttt{input2} at \autoref{line:main_input2} to steer later control flow, and then calls \texttt{internal\_cplus\_demangle} with both parameters at \autoref{line:main_call} to start demangling.
In this function, the code checks \texttt{flag} at \autoref{line:internal_cplus_demangle_flag} to choose between two branches, which we describe next:
\begin{enumerate}[noitemsep,topsep=0pt,parsep=0pt,partopsep=0pt]
    \item If \texttt{flag} is \texttt{true}, the code checks at \autoref{line:internal_cplus_demangle_if1} whether \texttt{mangled} is empty; if not, it parses the input, writes to the \texttt{output} buffer (allocated at \autoref{line:output}) at \autoref{line:internal_cplus_demangle_memcpy}, advances the buffer pointer at \autoref{line:internal_cplus_demangle_add1}, and calls \texttt{gnu\_special} at \autoref{line:internal_cplus_demangle_call1}.
    \item If \texttt{flag} is \texttt{false}, the code checks at \autoref{line:internal_cplus_demangle_if2} whether \texttt{mangled} starts with \texttt{\_GLOBAL\_} and the next character is not a digit; if so, it finishes parsing \texttt{\_GLOBAL\_}, advances \texttt{mangled} at \autoref{line:internal_cplus_demangle_add2}, and calls \texttt{gnu\_special} at \autoref{line:internal_cplus_demangle_call2}.
\end{enumerate}

\begin{table}[t!]
    \scriptsize
    \centering
    \caption{Information of two seeds $s_0$ and $s_1$.}
    \vspace{-0.1in}
    \label{seed-input}
    \begin{tabular}{ccc}
        \toprule
        \textbf{Seed}&\textbf{\texttt{input1}}&\textbf{\texttt{input2}} \\
        \midrule
        $s_0$ & \texttt{Alice} & \texttt{true} \\
        $s_1$ & \texttt{\_GLOBAL\_Bob} & \texttt{false} \\
        \bottomrule
    \end{tabular}
\end{table}

In \texttt{gnu\_special}, the code checks at \autoref{line:gnu_special_if} whether \texttt{mangled} starts with a digit; if so, it parses consecutive digits into \texttt{n} from \autoref{line:gnu_special_init} to \autoref{line:gnu_special_add1}, and then copies \texttt{n} characters from \texttt{mangled} to the output buffer at \autoref{line:gnu_special_cve}. Because \texttt{n} is not range-checked, it can become huge or overflow negative, causing a segmentation fault at \autoref{line:gnu_special_cve} and thus CVE-2016-4489.

DGF currently has two seeds, $s_0$ and $s_1$, and must decide which to mutate first. As shown in \autoref{seed-input}, neither seed can directly trigger the vulnerability because \texttt{input1} in both cases contains no digits. However, $s_0$ sets \texttt{input2} to \texttt{true} while $s_1$ sets it to \texttt{false}. If \texttt{input2} stays \texttt{false}, the vulnerability cannot be reached: \texttt{internal\_cplus\_demangle} takes its second branch, and the check in \autoref{line:internal_cplus_demangle_if2} ensures that the first character seen by \autoref{line:gnu_special_if} is not a digit, so the code at \autoref{line:gnu_special_cve} is unreachable. Therefore, $s_0$ deserves higher mutation priority: it only needs to mutate the start of \texttt{input1} into a large number, whereas $s_1$ must also flip \texttt{input2} to \texttt{true}. The next subsection shows how existing imprecise distance metrics fail on this seed prioritization task.

\subsection{Limitations of Existing Approaches with Imprecise Score Metrics}

We analyze three DGF approaches that rely on imprecise distance metrics, namely AFLGo \cite{AFLGo}, \textsc{WindRanger} \cite{Windranger}, and DAFL \cite{DAFL}, showing that they can mis-prioritize seeds in the above case. These methods derive distances from static analysis and use them for seed prioritization, so they may be misled by control-flow or data-flow relations that do not occur in real vulnerability-triggering executions. Although they use different metrics, all of them depend directly on static analysis, which cannot be fully accurate~\cite{Rice}.
Moreover, DGF typically compiles each software version separately, so heavy static analysis is too costly; many tools therefore use lightweight, coarse-grained analysis, which produces more imprecision and further aggravates the problem.

\begin{figure}[t!]
\centering
\begin{subfigure}[t]{0.49\textwidth}
\centering
\begin{adjustbox}{center,scale=0.7}
\begin{tikzpicture}[
    every node/.style={circle, draw}, node distance=0.3cm,font=\small
]
    \node[fill=lightgray, very thick] (31) {\ref{line:main_input1}};
    \node[below=of 31, fill=lightgray, very thick] (32) {\ref{line:main_input2}};
    \node[below=of 32, fill=lightgray, very thick] (33) {\underline{\ref{line:main_call}}};
    \node[below=of 33, fill=lightgray, very thick] (16) {\ref{line:internal_cplus_demangle_flag}};
    
    \node[below left=of 16, double, double distance = 2pt, fill=lightgray, very thick] (17) {\ref{line:internal_cplus_demangle_if1}};
    \node[below=of 17, fill=lightgray, very thick] (18) {\ref{line:internal_cplus_demangle_memcpy}};
    \node[draw=none, left=of 18] (17-DBB) {$\cdots$};
    \node[below=of 18, fill=lightgray, very thick] (19) {\ref{line:internal_cplus_demangle_add1}};
    \node[below=of 19, fill=lightgray, very thick] (20) {\underline{\ref{line:internal_cplus_demangle_call1}}};
    
    \node[below right=of 16, double, double distance = 2pt] (23) {\ref{line:internal_cplus_demangle_if2}};
    \node[below=of 23] (24) {\ref{line:internal_cplus_demangle_add2}};
    \node[draw=none, right=of 24] (23-DBB) {$\cdots$};
    \node[below=of 24] (25) {\ref{line:internal_cplus_demangle_call2}};
    
    \node[below right= of 20, double, double distance = 2pt, fill=lightgray, very thick] (4) {\ref{line:gnu_special_if}};
    \node[below=of 4, very thick] (5) {\ref{line:gnu_special_init}};
    \node[draw=none, left=of 5] (4-DBB) {$\cdots$};
    
    \node[below left=of 5, very thick] (6) {\ref{line:gnu_special_while}};
    \node[below right=of 6, very thick] (11) {\underline{\textbf{\ref{line:gnu_special_cve}}}};
    \node[below left=of 6, very thick] (7) {\ref{line:gnu_special_mul}};
    \node[below=of 7, very thick] (8) {\ref{line:gnu_special_add}};
    \node[below=of 8, very thick] (9) {\ref{line:gnu_special_add1}};

    \draw[->] (31) -- (32);
    \draw[->] (32) -- (33);
    \draw[->] (33) -- (16);
    \draw[->] (16) -- (17);
    \draw[->] (16) -- (23);
    \draw[->] (17) -- (18);
    \draw[->] (17) -- (17-DBB);
    \draw[->] (18) -- (19);
    \draw[->] (19) -- (20);
    \draw[->] (20) -- (4);
    \draw[->] (23) -- (24);
    \draw[->] (23) -- (23-DBB);
    \draw[->] (24) -- (25);
    \draw[->] (25) -- (4);
    \draw[->] (4) -- (5);
    \draw[->] (4) -- (4-DBB);
    \draw[->] (5) -- (6);
    \draw[->] (6) -- (11);
    \draw[->] (6) -- (7);
    \draw[->] (7) -- (8);
    \draw[->] (8) -- (9);
    \draw[->] (9) to[bend right] (6);

    \node[circle, fill=white, draw, inner sep=1pt, font=\scriptsize, fill=black, text=white] at (31.east) {10};
    \node[circle, fill=white, draw, inner sep=1.5pt, font=\scriptsize, fill=black, text=white] at (32.east) {9};
    \node[circle, fill=white, draw, inner sep=1.5pt, font=\scriptsize, fill=black, text=white] at (33.east) {8};
    \node[circle, fill=white, draw, inner sep=1.5pt, font=\scriptsize, fill=black, text=white] at (16.east) {7};
    \node[circle, fill=white, draw, inner sep=1.5pt, font=\scriptsize, fill=black, text=white] at (17.east) {7};
    \node[circle, fill=white, draw, inner sep=1.5pt, font=\scriptsize, fill=black, text=white] at (23.east) {6};
    \node[circle, fill=white, draw, inner sep=1.5pt, font=\scriptsize, fill=black, text=white] at (18.east) {6};
    \node[circle, fill=white, draw, inner sep=1.5pt, font=\scriptsize, fill=black, text=white] at (24.east) {5};
    \node[circle, fill=white, draw, inner sep=1.5pt, font=\scriptsize, fill=black, text=white] at (19.east) {5};
    \node[circle, fill=white, draw, inner sep=1.5pt, font=\scriptsize, fill=black, text=white] at (25.east) {4};
    \node[circle, fill=white, draw, inner sep=1.5pt, font=\scriptsize, fill=black, text=white] at (20.east) {4};
    \node[circle, fill=white, draw, inner sep=1.5pt, font=\scriptsize, fill=black, text=white] at (4.east) {3};
    \node[circle, fill=white, draw, inner sep=1.5pt, font=\scriptsize, fill=black, text=white] at (5.east) {2};
    \node[circle, fill=white, draw, inner sep=1.5pt, font=\scriptsize, fill=black, text=white] at (6.east) {1};
    \node[circle, fill=white, draw, inner sep=1.5pt, font=\scriptsize, fill=black, text=white] at (11.east) {0};
    \node[circle, fill=white, draw, inner sep=1.5pt, font=\scriptsize, fill=black, text=white] at (7.east) {4};
    \node[circle, fill=white, draw, inner sep=1.5pt, font=\scriptsize, fill=black, text=white] at (8.east) {3};
    \node[circle, fill=white, draw, inner sep=1.5pt, font=\scriptsize, fill=black, text=white] at (9.east) {2};

\end{tikzpicture}
\end{adjustbox}
\caption{The execution path of seed $s_0$.}
\end{subfigure}   
\begin{subfigure}[t]{0.49\textwidth}
\centering
\begin{adjustbox}{center,scale=0.7}
\begin{tikzpicture}[
    every node/.style={circle, draw}, node distance=0.3cm,font=\small
]
    \node[fill=lightgray, very thick] (31) {\ref{line:main_input1}};
    \node[below=of 31, fill=lightgray, very thick] (32) {\ref{line:main_input2}};
    \node[below=of 32, fill=lightgray, very thick] (33) {\underline{\ref{line:main_call}}};
    \node[below=of 33, fill=lightgray, very thick] (16) {\ref{line:internal_cplus_demangle_flag}};
    
    \node[below left=of 16, double, double distance = 2pt, very thick] (17) {\ref{line:internal_cplus_demangle_if1}};
    \node[below=of 17, very thick] (18) {\ref{line:internal_cplus_demangle_memcpy}};
    \node[draw=none, left=of 18] (17-DBB) {$\cdots$};
    \node[below=of 18, very thick] (19) {\ref{line:internal_cplus_demangle_add1}};
    \node[below=of 19, very thick] (20) {\underline{\ref{line:internal_cplus_demangle_call1}}};
    
    \node[below right=of 16, double, double distance = 2pt, fill=lightgray] (23) {\ref{line:internal_cplus_demangle_if2}};
    \node[below=of 23, fill=lightgray] (24) {\ref{line:internal_cplus_demangle_add2}};
    \node[draw=none, right=of 24] (23-DBB) {$\cdots$};
    \node[below=of 24, fill=lightgray] (25) {\ref{line:internal_cplus_demangle_call2}};
    
    \node[below right= of 20, double, double distance = 2pt, fill=lightgray, very thick] (4) {\ref{line:gnu_special_if}};
    \node[below=of 4, very thick] (5) {\ref{line:gnu_special_init}};
    \node[draw=none, left=of 5] (4-DBB) {$\cdots$};
    
    \node[below left=of 5, very thick] (6) {\ref{line:gnu_special_while}};
    \node[below right=of 6, very thick] (11) {\underline{\textbf{\ref{line:gnu_special_cve}}}};
    \node[below left=of 6, very thick] (7) {\ref{line:gnu_special_mul}};
    \node[below=of 7, very thick] (8) {\ref{line:gnu_special_add}};
    \node[below=of 8, very thick] (9) {\ref{line:gnu_special_add1}};

    \draw[->] (31) -- (32);
    \draw[->] (32) -- (33);
    \draw[->] (33) -- (16);
    \draw[->] (16) -- (17);
    \draw[->] (16) -- (23);
    \draw[->] (17) -- (18);
    \draw[->] (17) -- (17-DBB);
    \draw[->] (18) -- (19);
    \draw[->] (19) -- (20);
    \draw[->] (20) -- (4);
    \draw[->] (23) -- (24);
    \draw[->] (23) -- (23-DBB);
    \draw[->] (24) -- (25);
    \draw[->] (25) -- (4);
    \draw[->] (4) -- (5);
    \draw[->] (4) -- (4-DBB);
    \draw[->] (5) -- (6);
    \draw[->] (6) -- (11);
    \draw[->] (6) -- (7);
    \draw[->] (7) -- (8);
    \draw[->] (8) -- (9);
    \draw[->] (9) to[bend right] (6);

    \node[circle, fill=white, draw, inner sep=1pt, font=\scriptsize, fill=black, text=white] at (31.east) {10};
    \node[circle, fill=white, draw, inner sep=1.5pt, font=\scriptsize, fill=black, text=white] at (32.east) {9};
    \node[circle, fill=white, draw, inner sep=1.5pt, font=\scriptsize, fill=black, text=white] at (33.east) {8};
    \node[circle, fill=white, draw, inner sep=1.5pt, font=\scriptsize, fill=black, text=white] at (16.east) {7};
    \node[circle, fill=white, draw, inner sep=1.5pt, font=\scriptsize, fill=black, text=white] at (17.east) {7};
    \node[circle, fill=white, draw, inner sep=1.5pt, font=\scriptsize, fill=black, text=white] at (23.east) {6};
    \node[circle, fill=white, draw, inner sep=1.5pt, font=\scriptsize, fill=black, text=white] at (18.east) {6};
    \node[circle, fill=white, draw, inner sep=1.5pt, font=\scriptsize, fill=black, text=white] at (24.east) {5};
    \node[circle, fill=white, draw, inner sep=1.5pt, font=\scriptsize, fill=black, text=white] at (19.east) {5};
    \node[circle, fill=white, draw, inner sep=1.5pt, font=\scriptsize, fill=black, text=white] at (25.east) {4};
    \node[circle, fill=white, draw, inner sep=1.5pt, font=\scriptsize, fill=black, text=white] at (20.east) {4};
    \node[circle, fill=white, draw, inner sep=1.5pt, font=\scriptsize, fill=black, text=white] at (4.east) {3};
    \node[circle, fill=white, draw, inner sep=1.5pt, font=\scriptsize, fill=black, text=white] at (5.east) {2};
    \node[circle, fill=white, draw, inner sep=1.5pt, font=\scriptsize, fill=black, text=white] at (6.east) {1};
    \node[circle, fill=white, draw, inner sep=1.5pt, font=\scriptsize, fill=black, text=white] at (11.east) {0};
    \node[circle, fill=white, draw, inner sep=1.5pt, font=\scriptsize, fill=black, text=white] at (7.east) {4};
    \node[circle, fill=white, draw, inner sep=1.5pt, font=\scriptsize, fill=black, text=white] at (8.east) {3};
    \node[circle, fill=white, draw, inner sep=1.5pt, font=\scriptsize, fill=black, text=white] at (9.east) {2};

\end{tikzpicture}
\end{adjustbox}
\caption{The execution path of seed $s_1$.}
\end{subfigure}   
\vspace{-0.1in}
\caption{CFG corresponding to the code fragment in \autoref{code-cxxfilt} and the nodes traversed by the execution paths of seeds $s_0$ and $s_1$. For simplicity, each statement is treated as a separate basic block. The numbers within each node represent the line numbers in \autoref{code-cxxfilt}, while the numbers in the small circles to the right of each node indicate their shortest distances to the target node. Nodes with gray backgrounds represent the nodes traversed by the seed's execution path. Nodes with double borders represent deviation basic blocks (DBBs) as defined in \textsc{WindRanger}, which can reach the target node themselves but point to a node that cannot reach the target node. Nodes with bold numbers are target nodes. Nodes with bold borders belong to the ground-truth vulnerability-reaching control-flow trace, and nodes with underlined numbers belong to the ground-truth vulnerability-triggering call-stack trace.}
\label{cfg}
\end{figure}

We first show how \textsc{AFLGo}~\cite{AFLGo} handles this case. For each seed execution, \textsc{AFLGo} collects the traversed CFG nodes and computes the harmonic mean of their shortest-path distances to the target statement, as shown in \autoref{cfg}. For simplicity, we treat each statement as its own basic block. Seeds with smaller average distances are considered easier to mutate into vulnerability-triggering inputs and are therefore prioritized. Under this scoring rule, the scores of $s_0$ and $s_1$ are computed as follows:
$$
\footnotesize
\begin{aligned}
\textsc{Score}_\textsc{AFLGo}(s_0)
&= 9\left(
\overbrace{10^{-1}}^{\text{Node \ref{line:main_input1}}} +
\overbrace{9^{-1}}^{\text{Node \ref{line:main_input2}}} +
\overbrace{8^{-1}}^{\text{Node \ref{line:main_call}}} +
\overbrace{7^{-1}}^{\text{Node \ref{line:internal_cplus_demangle_flag}}} +
\overbrace{7^{-1}}^{\text{Node \ref{line:internal_cplus_demangle_if1}}} +
\overbrace{6^{-1}}^{\text{Node \ref{line:internal_cplus_demangle_memcpy}}} +
\overbrace{5^{-1}}^{\text{Node \ref{line:internal_cplus_demangle_add1}}} +
\overbrace{4^{-1}}^{\text{Node \ref{line:internal_cplus_demangle_call1}}} +
\overbrace{3^{-1}}^{\text{Node \ref{line:gnu_special_if}}}
\right)^{-1} \approx 5.75 \\
\textsc{Score}_\textsc{AFLGo}(s_1)
&= 8\left(
\overbrace{10^{-1}}^{\text{Node \ref{line:main_input1}}} +
\overbrace{9^{-1}}^{\text{Node \ref{line:main_input2}}} +
\overbrace{8^{-1}}^{\text{Node \ref{line:main_call}}} +
\overbrace{7^{-1}}^{\text{Node \ref{line:internal_cplus_demangle_flag}}} +
\overbrace{6^{-1}}^{\text{Node \ref{line:internal_cplus_demangle_if2}}} +
\overbrace{5^{-1}}^{\text{Node \ref{line:internal_cplus_demangle_add2}}} +
\overbrace{4^{-1}}^{\text{Node \ref{line:internal_cplus_demangle_call2}}} +
\overbrace{3^{-1}}^{\text{Node \ref{line:gnu_special_if}}}
\right)^{-1} \approx 5.64
\end{aligned}
$$
Because $\textsc{Score}_\textsc{AFLGo}(s_0) > \textsc{Score}_\textsc{AFLGo}(s_1)$, \textsc{AFLGo} would prioritize mutating $s_1$, which is incorrect and inefficient. This happens because the CFG includes control-flow edges that never appear in vulnerability-triggering executions. As a result, static analysis assigns shorter distances to nodes \ref{line:internal_cplus_demangle_if2}, \ref{line:internal_cplus_demangle_add2}, and \ref{line:internal_cplus_demangle_call2}, even though they cannot lead to the vulnerability, than to nodes \ref{line:internal_cplus_demangle_if1}, \ref{line:internal_cplus_demangle_memcpy}, and \ref{line:internal_cplus_demangle_add1}.

We next describe how \textsc{WindRanger}~\cite{Windranger} addresses this case. \textsc{WindRanger} introduces deviation basic blocks (DBBs), which are CFG nodes that can reach the target but have outgoing edges to nodes that cannot. DBBs are shown with double borders in \autoref{cfg}. \textsc{WindRanger} argues that averaging distances over all executed nodes, as in \textsc{AFLGo}, is diluted by many irrelevant nodes; it therefore computes distances only from executed DBBs to the target and prioritizes seeds with smaller arithmetic means. Under this scoring rule, the scores of $s_0$ and $s_1$ are computed as follows:
$$
\footnotesize
\begin{aligned}
\textsc{Score}_\textsc{WindRanger}(s_0)=
\frac{
\overbrace{7}^{\text{Node \ref{line:internal_cplus_demangle_if1}}} +
\overbrace{3}^{\text{Node \ref{line:gnu_special_if}}}}{2} = 5 \quad\quad \quad
\textsc{Score}_\textsc{WindRanger}(s_1)=
\frac{
\overbrace{6}^{\text{Node \ref{line:internal_cplus_demangle_if2}}} +
\overbrace{3}^{\text{Node \ref{line:gnu_special_if}}}}{2} = 4.5
\end{aligned}
$$
Because $\textsc{Score}_\textsc{WindRanger}(s_0) > \textsc{Score}_\textsc{WindRanger}(s_1)$, \textsc{WindRanger} would still prioritize mutating $s_1$, which is incorrect and inefficient. Although \textsc{WindRanger} reduces dilution from irrelevant nodes, it remains vulnerable to imprecise control-flow edges: DBB~\ref{line:internal_cplus_demangle_if2} cannot reach the target in vulnerability-triggering executions, but static analysis assigns it a smaller distance than DBB~\ref{line:internal_cplus_demangle_if1}, leading to the wrong choice.

Finally, we show how \textsc{DAFL}~\cite{DAFL} handles this case. \textsc{DAFL} argues that CFG-based control-flow feedback is too imprecise, and therefore uses a value-flow graph (VFG), whose edges represent potential data flows and thus exclude many useless edges compared to a CFG. For each seed execution, \textsc{DAFL} collects the traversed VFG nodes, as shown in \autoref{vfg}, and prioritizes seeds based on their shortest distances to the target statement. Let $L$ be the maximum distance from any node to the target ($L=3$ in our example). A node at distance $i$ receives score $L-i+1$, and \textsc{DAFL} prefers seeds with larger total scores over traversed nodes. Under this scoring rule, the scores of $s_0$ and $s_1$ are computed as follows:
$$
\footnotesize
\begin{aligned}
\textsc{Score}_\textsc{DAFL}(s_0)&=
\overbrace{L-3+1}^{\text{Node \ref{line:main_input1}}} +
\overbrace{L-2+1}^{\text{Node \ref{line:main_call}}} +
\overbrace{L-1+1}^{\text{Node \ref{line:internal_cplus_demangle_call1}}}
= 6 \\
\textsc{Score}_\textsc{DAFL}(s_1)&=
\overbrace{L-3+1}^{\text{Node \ref{line:main_input1}}} +
\overbrace{L-2+1}^{\text{Node \ref{line:main_call}}} +
\overbrace{L-2+1}^{\text{Node \ref{line:internal_cplus_demangle_add2}}} +
\overbrace{L-1+1}^{\text{Node \ref{line:internal_cplus_demangle_call2}}}
= 8
\end{aligned}
$$
Because $\textsc{Score}_\textsc{DAFL}(s_0) < \textsc{Score}_\textsc{DAFL}(s_1)$, \textsc{DAFL} would prioritize mutating $s_1$, making the same incorrect choice as \textsc{AFLGo} and \textsc{WindRanger}. Although the VFG removes many meaningless CFG edges, it is still produced by static analysis and thus remains imprecise, so it can include data flows that never occur in real vulnerability-triggering executions. In our case, nodes \ref{line:internal_cplus_demangle_add2} and \ref{line:internal_cplus_demangle_call2} are mistakenly treated as having data flows that can reach the target, which inflates the score of $s_1$.

In summary, we have demonstrated the limitations of three current DGF approaches that use imprecise distance metrics. To address this issue, we need to find more precise information beyond static analysis to guide DGF in seed prioritization.

\begin{figure}[t!]
\centering
\begin{subfigure}[t]{0.49\textwidth}
\centering
\begin{adjustbox}{center,scale=0.7}
\begin{tikzpicture}[
    every node/.style={circle, draw}, node distance=0.3cm,font=\small
]
    \node[fill=lightgray, very thick] (31) {\ref{line:main_input1}};
    \node[below=of 31, fill=lightgray, very thick] (33) {\ref{line:main_call}};
    
    \node[below left=of 33, fill=lightgray, very thick] (20) {\ref{line:internal_cplus_demangle_call1}};
    \node[below right=of 33] (24) {\ref{line:internal_cplus_demangle_add2}};
    \node[below =of 24] (25) {\ref{line:internal_cplus_demangle_call2}};
    \node[below =of 25] (9) {\ref{line:gnu_special_add1}};
    \node[below =of 9, very thick] (11) {\textbf{\ref{line:gnu_special_cve}}};

    \node[below =of 20, very thick] (5) {\ref{line:gnu_special_init}};
    \node[below =of 5, very thick] (7) {\ref{line:gnu_special_mul}};
    \node[below =of 7, very thick] (8) {\ref{line:gnu_special_add}};

    \draw[->] (31) -- (33);
    \draw[->] (33) -- (20);
    \draw[->] (33) -- (24);
    \draw[->] (24) -- (25);
    \draw[->] (25) -- (9);
    \draw[->] (20) to[bend left=6] (9);
    \draw[->] (20) to[bend left=5] (11);
    \draw[->] (25) to[bend left=50] (11);
    \draw[->] (9) -- (11);

    \draw[->] (5) -- (7);
    \draw[->] (7) -- (8);
    \draw[->] (20) to[bend right] (8);
    \draw[->] (25) to[bend right=8] (8);
    \draw[->] (9) to[bend right=13] (8);

    \draw[->] (8) -- (11);

    \node[circle, fill=white, draw, inner sep=1.5pt, font=\scriptsize, fill=black, text=white] at (11.east) {0};
    \node[circle, fill=white, draw, inner sep=1.5pt, font=\scriptsize, fill=black, text=white] at (8.east) {1};
    \node[circle, fill=white, draw, inner sep=1.5pt, font=\scriptsize, fill=black, text=white] at (9.east) {1};
    \node[circle, fill=white, draw, inner sep=1.5pt, font=\scriptsize, fill=black, text=white] at (20.east) {1};
    \node[circle, fill=white, draw, inner sep=1.5pt, font=\scriptsize, fill=black, text=white] at (25.east) {1};
    \node[circle, fill=white, draw, inner sep=1.5pt, font=\scriptsize, fill=black, text=white] at (7.east) {2};
    \node[circle, fill=white, draw, inner sep=1.5pt, font=\scriptsize, fill=black, text=white] at (24.east) {2};
    \node[circle, fill=white, draw, inner sep=1.5pt, font=\scriptsize, fill=black, text=white] at (33.east) {2};
    \node[circle, fill=white, draw, inner sep=1.5pt, font=\scriptsize, fill=black, text=white] at (31.east) {3};
    \node[circle, fill=white, draw, inner sep=1.5pt, font=\scriptsize, fill=black, text=white] at (5.east) {3};
\end{tikzpicture}
\end{adjustbox}
\caption{The execution path of seed $s_0$.}
\end{subfigure}   
\begin{subfigure}[t]{0.49\textwidth}
\centering
\begin{adjustbox}{center,scale=0.7}
\begin{tikzpicture}[
    every node/.style={circle, draw}, node distance=0.3cm,font=\small
]
    \node[fill=lightgray, very thick] (31) {\ref{line:main_input1}};
    \node[below=of 31, fill=lightgray, very thick] (33) {\ref{line:main_call}};
    
    \node[below left=of 33, very thick] (20) {\ref{line:internal_cplus_demangle_call1}};
    \node[below right=of 33, fill=lightgray] (24) {\ref{line:internal_cplus_demangle_add2}};
    \node[below =of 24, fill=lightgray] (25) {\ref{line:internal_cplus_demangle_call2}};
    \node[below =of 25] (9) {\ref{line:gnu_special_add1}};
    \node[below =of 9, very thick] (11) {\textbf{\ref{line:gnu_special_cve}}};

    \node[below =of 20, very thick] (5) {\ref{line:gnu_special_init}};
    \node[below =of 5, very thick] (7) {\ref{line:gnu_special_mul}};
    \node[below =of 7, very thick] (8) {\ref{line:gnu_special_add}};

    \draw[->] (31) -- (33);
    \draw[->] (33) -- (20);
    \draw[->] (33) -- (24);
    \draw[->] (24) -- (25);
    \draw[->] (25) -- (9);
    \draw[->] (20) to[bend left=6] (9);
    \draw[->] (20) to[bend left=5] (11);
    \draw[->] (25) to[bend left=50] (11);
    \draw[->] (9) -- (11);

    \draw[->] (5) -- (7);
    \draw[->] (7) -- (8);
    \draw[->] (20) to[bend right] (8);
    \draw[->] (25) to[bend right=8] (8);
    \draw[->] (9) to[bend right=13] (8);

    \draw[->] (8) -- (11);

    \node[circle, fill=white, draw, inner sep=1.5pt, font=\scriptsize, fill=black, text=white] at (11.east) {0};
    \node[circle, fill=white, draw, inner sep=1.5pt, font=\scriptsize, fill=black, text=white] at (8.east) {1};
    \node[circle, fill=white, draw, inner sep=1.5pt, font=\scriptsize, fill=black, text=white] at (9.east) {1};
    \node[circle, fill=white, draw, inner sep=1.5pt, font=\scriptsize, fill=black, text=white] at (20.east) {1};
    \node[circle, fill=white, draw, inner sep=1.5pt, font=\scriptsize, fill=black, text=white] at (25.east) {1};
    \node[circle, fill=white, draw, inner sep=1.5pt, font=\scriptsize, fill=black, text=white] at (7.east) {2};
    \node[circle, fill=white, draw, inner sep=1.5pt, font=\scriptsize, fill=black, text=white] at (24.east) {2};
    \node[circle, fill=white, draw, inner sep=1.5pt, font=\scriptsize, fill=black, text=white] at (33.east) {2};
    \node[circle, fill=white, draw, inner sep=1.5pt, font=\scriptsize, fill=black, text=white] at (31.east) {3};
    \node[circle, fill=white, draw, inner sep=1.5pt, font=\scriptsize, fill=black, text=white] at (5.east) {3};

\end{tikzpicture}
\end{adjustbox}
\caption{The execution path of seed $s_1$.}
\end{subfigure}   
\vspace{-0.1in}
\caption{VFG for the code in \autoref{code-cxxfilt}, highlighting the execution paths of seeds $s_0$ and $s_1$.
Node labels are line numbers in \autoref{code-cxxfilt}. The numbers in the small circles to the right of each node indicate their shortest distances to the target node.
Gray nodes are traversed by the seeds; bold numbers mark targets.
Bold-bordered nodes lie on the ground-truth vulnerability-reaching value-flow trace.}
\label{vfg}
\end{figure}

\subsection{Our Approach}
We propose trace-guided directed greybox fuzzing (TDGF), which uses execution information that triggers the target vulnerability as a guidance trace to prioritize seeds. By abandoning imprecise static-analysis-based distance metrics, TDGF effectively avoids the misleading guidance caused by over-approximation. TDGF can leverage any type of execution information collected at vulnerability triggering time; in this work, we empirically study two representative trace types: (1) fine-grained vulnerability-reaching control flow and (2) coarse-grained call stacks at the vulnerability-triggering point. We detail this empirical study in \autoref{section-4}. Below, we explain how TDGF instantiated with these two traces addresses the seed-prioritization problem in our case study.

We illustrate the two types of guidance traces in our case study:
$$
\footnotesize
\begin{aligned}
\textsc{Trace}_\text{ControlFlow}=
\{\ref{line:main_input1},\ref{line:main_input2},\ref{line:main_call},\ref{line:internal_cplus_demangle_flag},\ref{line:internal_cplus_demangle_if1},\ref{line:internal_cplus_demangle_memcpy},\ref{line:internal_cplus_demangle_add1},\ref{line:internal_cplus_demangle_call1},\ref{line:gnu_special_if},\ref{line:gnu_special_init},\ref{line:gnu_special_while},\ref{line:gnu_special_mul},\ref{line:gnu_special_add},\ref{line:gnu_special_add1},\ref{line:gnu_special_cve}\} \quad \quad \quad
\textsc{Trace}_\text{CallStack}=
\{\ref{line:main_call},\ref{line:internal_cplus_demangle_call1},\ref{line:gnu_special_cve}\} 
\end{aligned}
$$
We model each trace as a set because, for efficiency, greybox fuzzing~\cite{AFL} does not record the exact order or frequency of executed basic blocks; instead, it buckets this information. As a result, we can only tell whether a basic block was executed, so representing the guidance trace as a set makes the following computations more intuitive. In \autoref{cfg}, we draw nodes in $\textsc{Trace}_\text{ControlFlow}$ with bold borders and mark nodes in $\textsc{Trace}_\text{CallStack}$ by underlining their node numbers. $\textsc{Trace}_\text{ControlFlow}$ includes all basic blocks executed in the vulnerability-triggering run, whereas $\textsc{Trace}_\text{CallStack}$ includes the triggering basic block and the call sites on the call stack at that point.

TDGF computes, for each seed, the size of the intersection between its execution trace and the guidance trace. 
TDGF then prioritizes seeds with larger intersections for mutation. 
The intuition is that such seeds align more closely with the guidance trace and thus have a higher potential to mutate into inputs that trigger the target vulnerability. 
The results are as follows:
$$
\footnotesize
\begin{aligned}
&\textsc{Score}_{\textsc{Trace}_{\text{ControlFlow}}}(s_0)=
|\{\ref{line:main_input1},\ref{line:main_input2},\ref{line:main_call},\ref{line:internal_cplus_demangle_flag},\ref{line:internal_cplus_demangle_if1},\ref{line:internal_cplus_demangle_memcpy},\ref{line:internal_cplus_demangle_add1},\ref{line:internal_cplus_demangle_call1},\ref{line:gnu_special_if}\} \cap \textsc{Trace}_{\text{ControlFlow}}|=9 \\
&\textsc{Score}_{\textsc{Trace}_{\text{ControlFlow}}}(s_1)=
|\{\ref{line:main_input1},\ref{line:main_input2},\ref{line:main_call},\ref{line:internal_cplus_demangle_flag},\ref{line:internal_cplus_demangle_if2},\ref{line:internal_cplus_demangle_add2},\ref{line:internal_cplus_demangle_call2},\ref{line:gnu_special_if}\} \cap \textsc{Trace}_{\text{ControlFlow}}|=5 \\
&\textsc{Score}_{\textsc{Trace}_{\text{CallStack}}}(s_0)=
|\{\ref{line:main_input1},\ref{line:main_input2},\ref{line:main_call},\ref{line:internal_cplus_demangle_flag},\ref{line:internal_cplus_demangle_if1},\ref{line:internal_cplus_demangle_memcpy},\ref{line:internal_cplus_demangle_add1},\ref{line:internal_cplus_demangle_call1},\ref{line:gnu_special_if}\} \cap \textsc{Trace}_{\text{CallStack}}|=2 \\
&\textsc{Score}_{\textsc{Trace}_{\text{CallStack}}}(s_1)=
|\{\ref{line:main_input1},\ref{line:main_input2},\ref{line:main_call},\ref{line:internal_cplus_demangle_flag},\ref{line:internal_cplus_demangle_if2},\ref{line:internal_cplus_demangle_add2},\ref{line:internal_cplus_demangle_call2},\ref{line:gnu_special_if}\} \cap \textsc{Trace}_{\text{CallStack}}|=1
\end{aligned}
$$
Since $\textsc{Score}_{\textsc{Trace}_{\text{ControlFlow}}}(s_0)>\textsc{Score}_{\textsc{Trace}_{\text{ControlFlow}}}(s_1)$ and $\textsc{Score}_{\textsc{Trace}_{\text{CallStack}}}(s_0)>\textsc{Score}_{\textsc{Trace}_{\text{CallStack}}}(s_1)$, TDGF prioritizes mutating $s_0$ under both types of guidance traces. 
TDGF makes the correct choice because it is not misled by imprecise static-analysis information. 

To deploy TDGF in practice, we must choose a guidance-trace type. Fine-grained control-flow traces can intuitively provide stronger guidance than coarse-grained call stacks, but guidance traces are not available in advance; a practical option is to predict them with LLMs, and call-stack traces are plausibly easier for LLMs to predict than control-flow traces. We therefore need a systematic empirical study to understand this trade-off and make an informed design choice. 
We formalize \uline{\textbf{the proposed TDGF framework}} in \autoref{section-3}, then \uline{\textbf{empirically compare the two trace types}} in \autoref{section-4}, and finally propose \textsc{Staczzer} \uline{\textbf{based on TDGF and the study results}} in \autoref{section-5}.

\section{The Trace-Guided Directed Greybox Fuzzing Framework}
\label{section-3}

We first present some preliminaries, and then introduce our method.

\subsection{Preliminaries}
For the program to be fuzzed, we provide the following formal definitions for the relevant content.

\begin{definition}[program input]
We define $\mathbf{S}$ as the set of program inputs. Each $s \in \mathbf{S}$ represents a possible input to the program.
\end{definition}

\begin{definition}[basic block]
We define $\mathbf{B}$ as the set of basic blocks. Each $b \in \mathbf{B}$ represents a specific basic block in the program.
\end{definition}
Note that, in the example shown in \autoref{section-2}, we treat each statement as a separate basic block for simplicity. In our experiments, basic blocks follow the standard definition, i.e., a straight-line sequence of instructions.

\begin{definition}[crashing input]
A crashing input $s \in \mathbf{S}$ refers to an input that causes the program to crash at basic block $b \in \mathbf{B}$ when executed with this input. We use a function $\textsc{Crashing}(s,b) : \mathbf{S} \times \mathbf{B} \to \{\texttt{true},\texttt{false}\}$ to represent whether input $s$ causes the program to crash at basic block $b$.
\end{definition}

\begin{definition}[execution trace]
The execution trace of an input $s \in \mathbf{S}$ refers to all basic blocks traversed when the program runs with this input. We use a function $\textsc{Execution}(s) : \mathbf{S} \to \mathcal{P}(\mathbf{B})$ to represent the set.
\end{definition}
Here, an execution trace is defined as a set of basic blocks because, in greybox fuzzing \cite{AFL}, the order and frequency with which a seed executes basic blocks are coarsened by bucketing for efficiency. Under this design, we can only determine whether a seed has executed a given basic block. We follow this design throughout. Therefore, in this paper, we define a seed’s execution trace as the set of basic blocks it traverses.

\begin{definition}[goal of directed greybox fuzzing]
The goal of directed greybox fuzzing is to generate an input $s \in \mathbf{S}$ such that $\textsc{Crashing}(s,b_\text{target}) = \texttt{true}$ for a given target location $b_\text{target} \in \mathbf{B}$ in the shortest possible time.
\end{definition}

\subsection{Overview}

\autoref{TDGF} formalizes the workflow of trace-guided directed greybox fuzzing (TDGF), extending the greybox fuzzer AFL~\cite{AFL}. The two AFL-specific lines (\lineref{tdgf:choose} and \lineref{tdgf:assign}) modified in TDGF are highlighted in gray: (1) AFL's \textsc{ChooseSeed} selects a seed for prioritization; we instead prioritize seeds whose execution paths overlap more with the guidance traces. (2) AFL's \textsc{AssignEnergy} assigns mutation energy to the selected seed; we instead allocate more energy to seeds with higher overlap between their execution paths and the guidance traces. TDGF takes as input an initial seed set $S_0 \subseteq \mathbf{S}$, a target program location $b_\text{target} \in \mathbf{B}$, and a \textit{guidance trace} $\textsc{Guidance} \subseteq \mathbf{B}$. TDGF outputs the set of crashing inputs $S_\times$. TDGF maintains a current seed set $S$ (initialized to $S_0$) and a crashing set $S_\times$ (initialized to empty) (\lineref{tdgf:init}), and then repeatedly performs the following fuzzing loop (\lineref{tdgf:repeat}):

\begin{enumerate}
    \item TDGF selects a seed $s \in S$ to prioritize for mutation in the current round based on the guidance trace $\textsc{Guidance}$ (\lineref{tdgf:choose}, detailed in \autoref{section-3.3}), and assigns it an energy value $e \in \mathbb{N}$ based on $\textsc{Guidance}$ (\lineref{tdgf:assign}, detailed in \autoref{section-3.4}). Here, energy is the number of mutations performed in the round, following standard greybox fuzzing terminology~\cite{AFLFast,AFLGo,DAFL}. Assigning energy is commonly called power scheduling~\cite{AFLFast}; higher energy means higher priority and more mutation time.
    \item Given energy $e$, TDGF mutates seed $s$ for $e$ times (\lineref{tdgf:for}); each iteration calls $\textsc{Mutate}(s)$ to produce a new seed $s'$ (\lineref{tdgf:mutate}). $\textsc{Mutate}(s)$ uses the same mutation settings as \textsc{AFL}~\cite{AFL}, randomly choosing a mutator (for example, bit flips, byte-level add/subtract, deleting a random-length byte sequence, and so on).
    \item If the execution trace of $s'$ differs from that of all existing seeds (\lineref{tdgf:new}), TDGF keeps $s'$ (\lineref{tdgf:add1}); if $s'$ crashes (\lineref{tdgf:crash}), TDGF also adds it to $S_\times$ (\lineref{tdgf:add2}).
\end{enumerate}

When the time budget is exhausted (\lineref{tdgf:timeout}), TDGF returns the set of crashing seeds $S_\times$ (\lineref{tdgf:return}). TDGF aims to generate an $s'$ (\lineref{tdgf:mutate}) such that $\textsc{Crashing}(s', b_\text{target})=\texttt{true}$ as quickly as possible.

\begin{algorithm}[t!]
    \caption{Trace-guided directed greybox fuzzing.}
    \label{TDGF}
    \footnotesize
    \begin{algorithmic}[1]
        \Require{The set of initial seeds $S_0$, the target basic block $b_\text{target}$, and the guidance trace $\textsc{Guidance}$.}
        \Ensure{The crashing seeds set $S_\times$.}

        \Procedure{TDGF}{$S_0,b_\text{target},\textsc{Guidance}$}
            \State{$S \gets S_0, S_\times \gets \varnothing$}\label{tdgf:init}
            \Repeat\label{tdgf:repeat}
                \State{\colorbox{lightgray}{$s \gets \textsc{ChooseSeed}(S,\textsc{Guidance})$}}\label{tdgf:choose}
                \State{\colorbox{lightgray}{$e \gets \textsc{AssignEnergy}(S,s,\textsc{Guidance})$}}\label{tdgf:assign}
                \For{$i = 1 \to e$}\label{tdgf:for}
                    \State{$s' \gets \textsc{Mutate}(s)$}\label{tdgf:mutate}
                    \If{$\textsc{Execution}(s') \not \in \{\textsc{Execution}(s) \mid s \in S\}$}\label{tdgf:new}
                        \State{$S \gets S \cup \{s'\}$}\label{tdgf:add1}
                        \If{$\exists b \in \mathbf{B} \text{ such that } \textsc{Crashing}(s,b)=\texttt{true}$}\label{tdgf:crash}
                            \State{$S_\times \gets S_\times \cup \{s'\}$}\label{tdgf:add2}
                        \EndIf
                    \EndIf
                \EndFor
            \Until{\textit{timeout}}\label{tdgf:timeout}
            \State{\Return{$S_\times$}}\label{tdgf:return}
        \EndProcedure
    \end{algorithmic}
\end{algorithm}

\subsection{Seed Selection}\label{section-3.3}

\begin{algorithm}[t!]
    \caption{The algorithm to choose seed for mutation in TDGF.}
    \label{choose-seed}
    \footnotesize
    \begin{algorithmic}[1]
        \Require{The set of current seeds $S$, and the guidance trace $\textsc{Guidance}$.}
        \Ensure{The seed for mutation $s$.}

        \State{$S_\text{chosen} \gets \varnothing$}\label{choose-chosen}

        \Procedure{ChooseSeed}{$S,\textsc{Guidance}$}
            \If{$S_\text{chosen}=S$}\label{choose-empty}
                \State{$S_\text{chosen} \gets \varnothing$}\label{choose-clear}
            \EndIf
            \State{$S_\text{current} \gets S-S_\text{chosen}$}\label{choose-current}
            \State{$s \gets \arg\max_{s' \in S_\text{current}}\textsc{Score}\label{choose-score}(s',\textsc{Guidance})$}
            \State{$S_\text{chosen} \gets S_\text{chosen} \cup \{s\}$}\label{choose-add}
            \State{\Return{$s$}}\label{choose-return}
            
        \EndProcedure
    \end{algorithmic}
\end{algorithm}

\autoref{choose-seed} formalizes TDGF's seed-selection algorithm. The algorithm takes as input the current seed set $S \subseteq \mathbf{S}$ and the guidance trace $\textsc{Guidance}$. The algorithm outputs a seed $s \in S$ for priority mutation. TDGF maintains a global set $S_\text{chosen}$ of seeds already mutated in the current round (initialized to empty) (\lineref{choose-chosen}). If all seeds in $S$ have been mutated (\lineref{choose-empty}), TDGF starts a new round by clearing $S_\text{chosen}$ (\lineref{choose-clear}). TDGF then forms $S_\text{current}$ as the seeds not yet mutated in the current round (\lineref{choose-current}). For each $s' \in S_\text{current}$, TDGF computes the overlap score $\textsc{Score}(s',\textsc{Guidance})$ between the execution trace of $s'$ and $\textsc{Guidance}$ (\lineref{choose-score}), defined as follows:
$$
\small
\textsc{Score}(s',\textsc{Guidance})=|\textsc{Execution}(s') \cap \textsc{Guidance}|
$$
TDGF selects the seed $s$ with the largest overlap score for priority mutation, adds it to $S_\text{chosen}$ (\lineref{choose-add}), and returns $s$ (\lineref{choose-return}).

\subsection{Power Scheduling}\label{section-3.4}

\begin{algorithm}[t!]
    \caption{The algorithm to assign energy for the seed for mutation in TDGF.}
    \label{assign-energy}
    \footnotesize
    \begin{algorithmic}[1]
        \Require{The set of current seeds $S$, the seed for mutation $s$, and the guidance trace $\textsc{Guidance}$.}
        \Ensure{The energy assigned to the seed for mutation $e$.}

        \Procedure{AssignEnergy}{$S,s,\textsc{Guidance}$}
            \State{$\textsc{Score}_\text{average} \gets \dfrac{\sum_{s' \in S}\textsc{Score}(s',\textsc{Guidance})}{|S|}$}\label{assign-average}
            
            \State{$e \gets \left\lceil\dfrac{\textsc{Score}(s,\textsc{Guidance})}{\textsc{Score}_\text{average}} \cdot \textsc{Energy}_\textsc{AFL}(s)\right\rceil$}\label{assign-cal}
            \State{\Return{$e$}}\label{assign-return}
            
        \EndProcedure
    \end{algorithmic}
\end{algorithm}

\autoref{assign-energy} formalizes TDGF's energy-assignment algorithm. The algorithm takes as input the current seed set $S \subseteq \mathbf{S}$, the selected seed $s \in S$, and the guidance trace $\textsc{Guidance}$. The algorithm outputs an energy value $e \in \mathbb{N}$ for $s$. TDGF first computes the average overlap score $\textsc{Score}_\text{average}$ between the execution traces of all current seeds and $\textsc{Guidance}$ (\lineref{assign-average}). It then scales the baseline energy $\textsc{Energy}_\textsc{AFL}(s)$ that \textsc{AFL}~\cite{AFL} would assign to $s$ by an adjustment factor (\lineref{assign-cal}), defined as the ratio between $s$'s overlap score and $\textsc{Score}_\text{average}$. Thus, seeds with higher overlap receive more energy. Here, $\textsc{Energy}_\textsc{AFL}(s)$ is \textsc{AFL}'s heuristic energy based on metrics such as execution time and trace length. TDGF finally returns the assigned energy $e$ (\lineref{assign-return}).

\section{Empirical Study}\label{section-4}

We empirically study two typical guidance traces observed when triggering a vulnerability, namely the control-flow trace and the call-stack trace, denoted as \textsc{ControlFlow} and \textsc{CallStack}, respectively. We expect that the results of this study will help us make an informed design choice when incorporating TDGF into our final framework, thereby improving the effectiveness of DGF in practical settings.

In particular, we aim to answer the following three questions:
\begin{enumerate}[noitemsep,topsep=0pt,parsep=0pt,partopsep=0pt]
  \item How effective is TDGF when provided with ground-truth guidance traces?
  \item As a fine-grained guidance trace, how much improvement does \textsc{ControlFlow} offer compared to the coarse-grained \textsc{CallStack}?
  \item In practical engineering scenarios, how accurate are LLMs at predicting these two types of guidance traces?
\end{enumerate}

We first describe the experimental setup in \autoref{section-4.1}. We then answer the first two questions through experiments and analysis in \autoref{section-4.2}. Finally, we address the third question in \autoref{section-4.3}.

\subsection{Experimental Setup}\label{section-4.1}

We run all experiments on Linux machines with 256 processors (2.25\,GHz) and 256\,GB RAM, and implement TDGF on top of \textsc{AFL}~\cite{AFL}.

\textit{Baselines.} We compare TDGF with: (1) \textsc{AFLGo}~\cite{AFLGo}, which prioritizes seeds using CFG distances between execution traces and targets; (2) \textsc{WindRanger}~\cite{Windranger}, which refines \textsc{AFLGo} by computing distances only from deviation basic blocks to reduce interference from irrelevant blocks; (3) \textsc{DAFL}~\cite{DAFL}, a SOTA directed greybox fuzzer that prioritizes seeds using distances on a value-flow graph; and (4) \textsc{AFL}~\cite{AFL}, the base fuzzer for both our approach and the other baselines.

\begin{table}[t!]
\centering
\caption{Benchmarks used in the experiments. The \textbf{Type} column indicates the abbreviated type of each vulnerability: \textbf{BF} (Buffer Overflow), \textbf{NPD} (Null Pointer Dereference), \textbf{SO} (Stack Overflow), \textbf{UAF} (Use-After-Free), \textbf{IO} (Integer Overflow), \textbf{OOM} (Out-Of-Memory).}
\vspace{-0.1in}
\label{benchmark}
\tiny
\begin{subtable}[t]{0.50\textwidth}
\centering
\begin{tabular}[t]{lllcc}
\toprule
\textbf{Project}  & \textbf{Program}  & \textbf{Version} & \textbf{CVE ID}        & \textbf{Type} \\ \midrule
\multirow{17}{*}{Ming} & \multirow{17}{*}{\texttt{swftophp}}
 & \multirow{7}{*}{0.4.7} & 2016-9827  & BO \\*
 &        & & 2016-9829  & NPD  \\*
 &        & & 2016-9831  & BO \\*
 &        & & 2017-9988  & NPD  \\*
 &        & & 2017-11728 & NPD  \\*
 &        & & 2017-11729 & NPD  \\*
 &        & & 2017-7578  & BO \\ \cmidrule(l){3-5}
 &        & \multirow{10}{*}{0.4.8} & 2018-7868  & UAF \\*
 &        & & 2018-8807  & UAF \\*
 &        & & 2018-8962  & UAF \\*
 &        & & 2018-11095 & UAF \\*
 &        & & 2018-11225 & BO \\*
 &        & & 2018-11226 & BO \\*
 &        & & 2018-20427 & NPD  \\*
 &        & & 2019-9114  & BO \\*
 &        & & 2019-12982 & BO \\*
 &        & & 2020-6628  & BO \\ \midrule
\multirow{2}{*}{Lrzip} & \multirow{2}{*}{\texttt{lrzip}}
 & \multirow{2}{*}{0.631} & 2017-8846 & UAF \\*
 &        &        & 2018-11496 & UAF \\ 
 \midrule
\multirow{2}{*}{Binutils} & \multirow{2}{*}{\texttt{cxxfilt}}
 & \multirow{2}{*}{2.26} & 2016-4487 & NPD  \\*
 &     &     & 2016-4489 & UAF    \\*\bottomrule
\end{tabular}
\end{subtable}
\begin{subtable}[t]{0.49\textwidth}
\centering
\begin{tabular}[t]{lllcc}
\toprule
\textbf{Project}  & \textbf{Program}  & \textbf{Version} & \textbf{CVE ID}        & \textbf{Type} \\ \midrule
\multirow{15}{*}{Binutils} & \multirow{4}{*}{\texttt{cxxfilt}}
 & \multirow{4}{*}{2.26} & 2016-4490 & IO  \\*
 &     &     & 2016-4491 & SO  \\*
 &     &     & 2016-4492 & IO  \\*
 &     &     & 2016-6131 & SO  \\ \cmidrule(l){2-5}
 & \multirow{3}{*}{\texttt{objcopy}}
 & \multirow{3}{*}{2.28} & 2017-8393 & BO \\*
 &     &     & 2017-8394 & NPD  \\*
 &     &     & 2017-8395 & NPD  \\ \cmidrule(l){2-5}
 & \multirow{5}{*}{\texttt{objdump}}
 & \multirow{4}{*}{2.28} & 2017-8392 & NPD  \\*
 &     &     & 2017-8396 & BO \\*
 &     &     & 2017-8397 & BO \\*
 &     &     & 2017-8398 & BO \\ \cmidrule(l){3-5}
 &  & 2.31.1 & 2018-17360 & BO \\ \cmidrule(l){2-5}
 & \texttt{strip}   & 2.27 & 2017-7303  & NPD  \\*
 & \texttt{nm}      & 2.29 & 2017-14940 & OOM \\*
 & \texttt{readelf} & 2.29 & 2017-16828 & IO  \\ \midrule
\multirow{3}{*}{Libxml2} & \multirow{3}{*}{\texttt{xmllint}}
 & \multirow{3}{*}{2.9.4} & 2017-5969 & NPD  \\*
 &        &        & 2017-9047 & BO \\*
 &        &        & 2017-9048 & BO \\ \midrule
\multirow{2}{*}{Libjpeg} & \multirow{2}{*}{\texttt{cjpeg}}
 & 1.5.90 & 2018-14498 & BO \\*
 &        & 2.0.4  & 2020-13790 & BO \\ \bottomrule
\end{tabular}
\end{subtable}
\end{table}

\textit{Benchmarks.} \autoref{benchmark} lists our benchmarks, which include 41 known vulnerabilities from 10 programs. For each vulnerability, we report its CVE ID and type. We use the same benchmark as the SOTA DGF tool \textsc{DAFL}~\cite{DAFL}. These vulnerabilities come from programs widely used in prior DGF work~\cite{DAFL,Beacon,Hawkeye}, so the benchmark spans diverse scenarios and provides a representative testbed for evaluating DGF techniques. We compile each program with AddressSanitizer~\cite{ASan}, which reports a crash backtrace and lets us reliably determine whether an input triggers the target vulnerability.

\textit{Metrics.} For each fuzzer and vulnerability, we run 40 trials to mitigate randomness from mutation and set a 24-hour time limit per trial, matching \textsc{DAFL}~\cite{DAFL}. Each trial runs in a dedicated Docker container pinned to a dedicated CPU core. We record time-to-exposure (TTE), defined as the time until the first trigger of the vulnerability. Following \textsc{DAFL}, we use the median of the 40 TTEs. If at least 20 trials do not trigger within the time limit, the median is undefined, and we set it to a penalty value of twice the time limit (48\,h). We say a vulnerability is stably reproducible by a fuzzer if and only if its median TTE is defined without the penalty.
Consistent with \textsc{DAFL}, we include static-analysis overhead in TTE because directed fuzzing is often used for patch testing where each code change requires recompilation, so static analysis is not a one-time cost.

\textit{Guidance traces.} For each vulnerability, we use its proof-of-concept (PoC) input to trigger a crash, record the executed basic-block set as $\textsc{Guidance}_\textsc{ControlFlow}$, and collect the crash backtrace to build $\textsc{Guidance}_\textsc{CallStack}$ as the set of basic blocks on the reported call stack, including the basic block containing the crashing statement.

\subsection{Effectiveness of TDGF with Ground-Truth Guidance Traces}\label{section-4.2}

\begin{table}[t!]
    \centering
    \tiny
    \caption{Summary of results for TDGF with two types of guidance metrics and each baseline over 40 runs on 41 known vulnerabilities. The \textbf{$\text{TDGF}_T$’s average speedup} row ($T\in\{\textsc{ControlFlow}, \textsc{CallStack}\}$) denotes the average speedup (in terms of median TTE) achieved by $\text{TDGF}_T$ over the corresponding configuration across the 41 vulnerabilities. The \textbf{\# Stably reproducible} row reports the number of vulnerabilities for which each fuzzer can compute a valid median TTE. In the \textbf{Performance vs. $\textsc{TDGF}_T$} row, $a/b$ indicates that $\textsc{TDGF}_T$ outperforms the fuzzer on $a$ vulnerabilities while being outperformed on $b$ vulnerabilities. This value is used to calculate the sign test $p$-value shown in the next row.}
    \label{optimal}
    \vspace{-0.1in}
    \begin{tabular}{ccccccc}
    \toprule
    \multicolumn{1}{c}{\textbf{Metric}} &
    \multicolumn{1}{c}{\textbf{$\textsc{TDGF}_\textsc{ControlFlow}$}} &
    \multicolumn{1}{c}{\textbf{$\textsc{TDGF}_\textsc{CallStack}$}} &
    \multicolumn{1}{c}{\textbf{\textsc{AFLGo}}} &
    \multicolumn{1}{c}{\textbf{\textsc{WindRanger}}} &
    \multicolumn{1}{c}{\textbf{\textsc{DAFL}}} &
    \multicolumn{1}{c}{\textbf{\textsc{AFL}}} \\
    \midrule
    \textbf{$\textsc{TDGF}_\textsc{ControlFlow}$'s average speedup}&-&1.23&3.47&2.99&2.08&2.47 \\
    \textbf{$\textsc{TDGF}_\textsc{CallStack}$'s average speedup}&-&-&3.30&2.51&2.14&2.16 \\
    \textbf{\# Stably reproducible}&28&29&22&21&27&23 \\
    \textbf{\textbf{Performance vs. $\textsc{TDGF}_\textsc{ControlFlow}$}}&-&\textbf{20}/16&\textbf{32}/4&\textbf{27}/9&\textbf{28}/10&\textbf{25}/11 \\
    \textbf{P-value in the sign test }&-&0.31&$10^{-6}$&0.002&0.003&0.01 \\
    
    \textbf{\textbf{Performance vs. $\textsc{TDGF}_\text{CallStack}$}}&-&-&\textbf{30}/4&\textbf{29}/6&\textbf{28}/7&\textbf{27}/9 \\
    \textbf{P-value in the sign test }&-&-&$3\times10^{-6}$&$6\times10^{-5}$&$3\times10^{-4}$&0.002 \\
    \bottomrule
\end{tabular}
\end{table}

We denote the TDGF configuration that uses $\textsc{Guidance}_\textsc{ControlFlow}$ and $\textsc{Guidance}_\textsc{CallStack}$ as guidance traces by $\textsc{TDGF}_\textsc{ControlFlow}$ and $\textsc{TDGF}_\textsc{CallStack}$, respectively. \autoref{optimal} summarizes the results of $\textsc{TDGF}_T$ ($T\in\{\textsc{ControlFlow}, \textsc{CallStack}\}$) and several baselines on 41 known vulnerabilities.

To answer question~(1), we report the average speedup of \textsc{TDGF}$_T$ over the baselines. \textsc{TDGF}$_\textsc{ControlFlow}$ achieves $2.08\times$ to $3.47\times$ speedups, and \textsc{TDGF}$_\textsc{CallStack}$ achieves $2.14\times$ to $3.30\times$. They stably reproduce 28 and 29 vulnerabilities, respectively, exceeding all baselines.
We also report per-vulnerability win-loss results in the \textbf{Performance vs. \textsc{TDGF}$_T$} rows. For each vulnerability, we compare median TTE; if the median TTE is unavailable for both, we compare the number of successful triggers. Each result is written as $a/b$, where $a$ is the number of cases where \textsc{TDGF}$_T$ is better and $b$ is the number of cases where the baseline is better. The win-loss ratios range from $2.27\times$ to $8\times$ for \textsc{TDGF}$_\textsc{ControlFlow}$ and from $3\times$ to $7.5\times$ for \textsc{TDGF}$_\textsc{CallStack}$, highlighting the benefit of trace-guided fuzzing.
Finally, we perform a one-tailed sign test~\cite{SignTest}, where $a$ and $b$ are the numbers of positive and negative signs, respectively; the p-values (shown in the subsequent \textbf{P-value in the sign test} rows) are all far below 0.05. \textbf{\uline{Therefore, our answer to question (1) is that \textsc{TDGF}$_T$ significantly outperforms all baselines, and TDGF is effective with either type of ground-truth guidance trace.}}

\begin{figure}[t!]
    \centering
    \includegraphics[width=\textwidth]{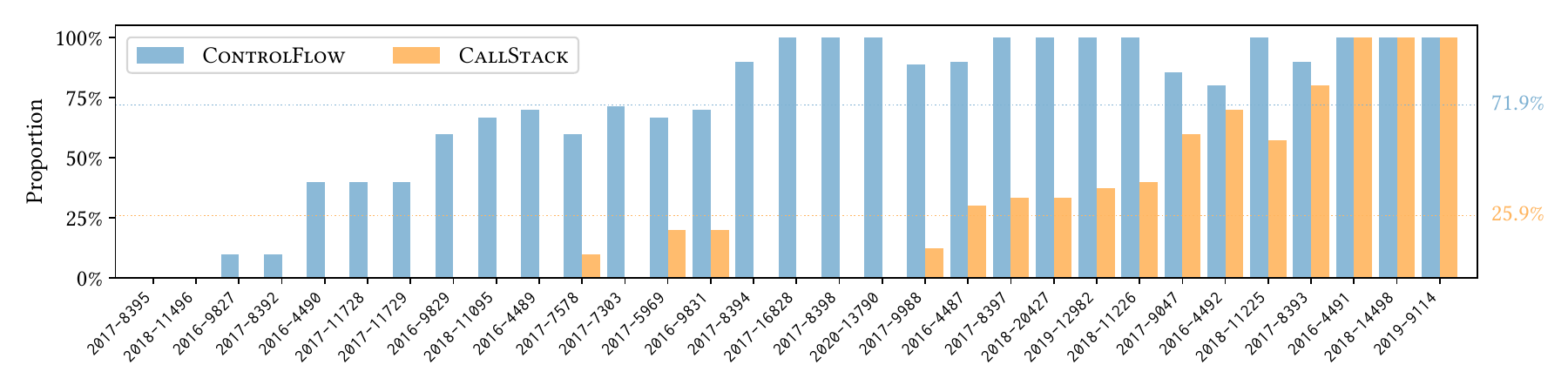}
    \vspace{-0.35in}
    \caption{The proportion of triggering-seed mutation paths whose scores are not monotonically non-decreasing. The dotted line represents the average value.}
    \label{study}
\end{figure}

To answer question~(2), we compare $\textsc{TDGF}_\textsc{ControlFlow}$ with $\textsc{TDGF}_\textsc{CallStack}$. $\textsc{TDGF}_\textsc{ControlFlow}$ is only $1.23\times$ faster and stably reproduces one fewer vulnerability. In the win-loss comparison, it wins on 20 vulnerabilities and loses on 16, so neither trace type consistently dominates. Accordingly, the one-tailed sign test fails to reject the null hypothesis, and we cannot claim that $\textsc{TDGF}_\textsc{ControlFlow}$ significantly outperforms $\textsc{TDGF}_\textsc{CallStack}$. Thus, although \textsc{ControlFlow} is more fine-grained, it does not provide significantly stronger guidance than the coarser \textsc{CallStack}. To explain this seemingly counter-intuitive result, we introduce triggering-seed mutation paths.
\begin{definition}[Triggering-seed mutation path]
Let $s_\text{trigger}$ be the earliest input generated by the fuzzer such that
$\textsc{Crash}(s_\text{trigger}, b_\text{target})=\texttt{true}$.
Then there exists a unique triggering-seed mutation path $(s_1,s_2,\dots,s_n)$ such that
$s_1\in S_0$, $s_n=s_\text{trigger}$, and for all $i\in\{2,\dots,n\}$, $s_i$ is mutated from $s_{i-1}$.
\end{definition}
Ideally, along a triggering-seed mutation path $(s_1,s_2,\dots,s_n)$, the TDGF scores
$\textsc{Score}(s_1,\textsc{Guidance}_T),$\allowbreak$ \textsc{Score}(s_2,\textsc{Guidance}_T), \dots,
\textsc{Score}(s_n,\textsc{Guidance}_T)$
should be monotonically non-decreasing, meaning that seeds move progressively closer to
the target trace.
\autoref{study} reports, across all runs of $\textsc{TDGF}_\textsc{ControlFlow}$ and
$\textsc{TDGF}_\textsc{CallStack}$, the fraction of triggering-seed mutation paths whose
score sequences are \textbf{not} monotonically non-decreasing. For \textsc{ControlFlow},
the average reaches 71.9\%, while for \textsc{CallStack} it is 25.9\%. This suggests that
under $\textsc{TDGF}_\textsc{ControlFlow}$, seeds off the triggering path often receive
higher scores than seeds on the path, whereas this happens much less frequently under
$\textsc{TDGF}_\textsc{CallStack}$.
A plausible explanation is that \textsc{ControlFlow} is too fine-grained: a single
mutation may reduce control-flow overlap before increasing it again later, so progress
is not smooth. The coarser \textsc{CallStack} trace mitigates this effect because one
mutation is less likely to change the call stack drastically. Consequently, the extra
fine-grained information in \textsc{ControlFlow} often behaves like noise: seeds that
should be prioritized typically have higher call-stack overlap, but do not necessarily
show higher control-flow overlap beyond that. Hence, \textsc{ControlFlow} offers limited
benefit over \textsc{CallStack} for guidance.
\textbf{\uline{Therefore, our answer to question~(2) is that \textsc{ControlFlow} does not provide
significantly stronger guidance than \textsc{CallStack}, due to its overly fine-grained nature.}}

\subsection{Accuracy of Guidance Trace Prediction in Practice}\label{section-4.3}

To answer question (3), we use two state-of-the-art code agents, Codex 0.75.0 \cite{codex} (GPT-5.1-codex-max high \cite{gpt-5.1-codex}) and Claude Code 2.0.76 \cite{claude-code} (Claude-Opus-4.5-20251101 \cite{claude-opus-4-5}), to predict the two types of guidance traces. We set the temperature to the default value of 1.0. \autoref{prompt} shows the prompt templates used; we query the code agents directly on each program's codebase. For evaluation, we corrected the ground truth by removing line numbers that do not exist in the codebase (e.g., those from dynamically linked libraries). \autoref{cfg-acc} reports the average accuracy under different metrics. Both code agents achieve very low accuracy on \textsc{ControlFlow}: all three metrics are around 0.2. In contrast, the metrics for \textsc{CallStack} are all above 0.7. This aligns with intuition: \textsc{CallStack} is more general and coarse-grained, which LLMs tend to handle well, whereas \textsc{ControlFlow} is fine-grained and requires rigorous logical reasoning, making it much harder for LLMs. \textbf{\uline{Therefore, in response to question (3), we conclude that \textsc{CallStack} is easier to predict than \textsc{ControlFlow}.}}

Based on our answers to the three questions, we conclude that \textsc{CallStack} provides nearly the same guidance capability as \textsc{ControlFlow}, yet is substantially easier for LLMs to predict, making it a more suitable guidance trace for TDGF. Motivated by this observation, we propose \textsc{Staczzer}, a framework that guides directed greybox fuzzing using LLM-predicted vulnerability-triggering call stacks. We introduce \textsc{Staczzer} in the next section.

\begin{figure}[t!]
\centering
\begin{tcolorbox}[
    colback=gray!20,
    colframe=darkgray,
    coltitle=white,
    colbacktitle=darkgray,
]
\scriptsize
\textbf{System:} You are a fuzzing expert, well-versed in the operation logic of AFL and its extension tools. You have a deep understanding of bugs inherent in the code.

\rule{\linewidth}{0.5pt}

\noindent
\begin{minipage}[t]{0.48\linewidth}
\textbf{Prompt Content for \textsc{ControlFlow}:}\\
In the current codebase, there may exist an input originating from the \texttt{<MAIN>} entry point that triggers a vulnerability in \texttt{<TARGET>}. Please identify the corresponding execution path and output every line of code executed along this path based on your code understanding and static reasoning.
\end{minipage}
\hfill
\vrule width 0.4pt
\hfill
\begin{minipage}[t]{0.48\linewidth}
\textbf{Prompt Content for \textsc{CallStack}:}\\
In the current codebase, there may exist an input originating from the \texttt{<MAIN>} entry point that triggers a vulnerability in \texttt{<TARGET>}. Please output the possible call stack that would directly trigger the crash (that is, if there were a sanitizer report) based on your code understanding and static reasoning.
\end{minipage}
\end{tcolorbox}
\vspace{-0.15in}
\caption{Prompt template for instructing the LLM to predict guidance traces. For brevity, formatting details and examples are omitted.}
\label{prompt}
\end{figure}

\begin{table}[t!]
\scriptsize
\centering
\caption{Average accuracy of guidance traces prediction using the code agents.}
\vspace{-0.1in}
\label{cfg-acc}
\begin{tabular}{lrrrrrr}
\toprule
\multicolumn{1}{c}{\multirow{2}{*}{\textbf{Type}}} &
\multicolumn{3}{c}{\textbf{Codex}} &
\multicolumn{3}{c}{\textbf{Claude Code}} \\
\cmidrule(lr){2-4}\cmidrule(lr){5-7}
& \multicolumn{1}{c}{\textbf{Precision}} & \multicolumn{1}{c}{\textbf{Recall}} & \multicolumn{1}{c}{\textbf{F1-Score}}
& \multicolumn{1}{c}{\textbf{Precision}} & \multicolumn{1}{c}{\textbf{Recall}} & \multicolumn{1}{c}{\textbf{F1-Score}} \\
\midrule
\textsc{ControlFlow} & 0.272 & 0.248 & 0.209 & 0.242 & 0.174 & 0.174 \\ \textsc{CallStack} & 0.766 & 0.719 & 0.730 & 0.748 & 0.709 & 0.718 \\
\bottomrule
\end{tabular}
\end{table}
\section{The \textsc{Staczzer} Framework}
\label{section-5}

\begin{algorithm}[t!]
    \caption{The \textsc{Staczzer} framework.}
    \label{staczzer}
    \footnotesize
    \begin{algorithmic}[1]
        \Require{The set of initial seeds $S_0$ and the target basic block $b_\text{target}$.}
        \Ensure{The crashing seeds set $S_\times$.}

        \Procedure{Staczzer}{$S_0,b_\text{target}$}
            \State{$\textsc{Slice} \gets \text{All function slices reaching the function containing $b_\text{target}$ on the call graph}$}\label{staczzer:cg}
            \If{$\text{Token count of \textsc{Slice} } \le 10^5$}\label{staczzer:if}
                \State{$\textsc{Guidance}_\textsc{PredictedCallStack} \gets \textsc{SingleTurnPrediction}()$}\label{staczzer:single}
            \Else
                \State{$\textsc{Guidance}_\textsc{PredictedCallStack} \gets \textsc{CodeAgentPrediction}()$}\label{staczzer:agent}
            \EndIf
            \State{\Return{$\textsc{TDGF}(S_0,b_\text{target},\textsc{Guidance}_\textsc{PredictedCallStack})$}}\label{staczzer:return}
        \EndProcedure
    \end{algorithmic}
\end{algorithm}

\autoref{staczzer} formalizes the workflow of the \textsc{Staczzer} framework. \textsc{Staczzer} follows the TDGF paradigm and uses LLM-predicted call stacks as the guidance trace. Given an initial seed set $S_0$ and a target program location $b_\text{target}$, \textsc{Staczzer} outputs the set of inputs that can trigger program crashes, denoted as $S_\times$.

In practice, DGF is typically applied to a specific software version; each code update requires recompilation and reprocessing, so the preprocessing cost is not a one-time overhead~\cite{DAFL}. With this in mind, \textsc{Staczzer} first performs static analysis to build the call graph, and then collects function-level slices for all functions that can reach the target function in the call graph (\lineref{staczzer:cg}). If the slice is small enough in terms of token count (\lineref{staczzer:if}), \textsc{Staczzer} directly feeds it to an LLM in a single turn to predict call stacks (\lineref{staczzer:single}); otherwise, \textsc{Staczzer} uses a code agent to perform the prediction (\lineref{staczzer:agent}). This design has two benefits:
\begin{enumerate}[noitemsep,topsep=0pt,parsep=0pt,partopsep=0pt]
  \item For small programs, code agents can be overly time-consuming. For some vulnerabilities, the cost can even exceed fuzzing itself. When the function slice is sufficiently small, single-turn LLM prediction makes the LLM overhead negligible. \textsc{Staczzer} sets the token threshold to $10^5$ to ensure prediction accuracy.
  \item For large programs, the relative overhead of using a code agent becomes less significant. Thus, we use code agents to improve prediction accuracy while maintaining scalability to large codebases.
\end{enumerate}
\textsc{Staczzer} uses the same prompt template as the one shown on the right side of \autoref{prompt}. For single-turn LLM prediction, \textsc{Staczzer} additionally includes the function slices in the prompt.
Finally, \textsc{Staczzer} uses the predicted $\textsc{Guidance}_\textsc{PredictedCallStack}$ as the guidance trace to run TDGF and returns the final results (\lineref{staczzer:return}).

\section{Experimental Evaluation}

Our evaluation aims to answer the following questions:

\begin{itemize}[noitemsep,topsep=0pt,parsep=0pt,partopsep=0pt]
    \item[\textbf{RQ1.}] How accurate is \textsc{Staczzer} in predicting call stacks? How effective is fuzzing when guided by the predicted call stacks?
    \item[\textbf{RQ2.}] Does the use of different LLMs affect the performance of \textsc{Staczzer}?
    \item[\textbf{RQ3.}] Do different components (seed selection, power scheduling, and LLM prediction) each contribute to the effectiveness of \textsc{Staczzer}?
    \item[\textbf{RQ4.}] Can \textsc{Staczzer} discover new real-world vulnerabilities?
\end{itemize}

We first describe our experimental setup in \autoref{section-6.1}. Then, we answer the four research questions in \autoref{section-6.2} to \autoref{section-6.5}, respectively.

\subsection{Experimental Setup}\label{section-6.1}

Our experimental setup is based on \autoref{section-4.1} with several improvements. Here we describe the additional settings; all unspecified settings remain exactly the same as in \autoref{section-4.1}. 

We use SVF~\cite{SVF} for static analysis. For \textsc{SingleTurnPrediction}, we use Claude-Opus-4.5-20251101. For \textsc{CodeAgentPrediction}, we use Claude Code~2.0.76 \cite{claude-code} (Claude-Opus-4.5-20251101 \cite{claude-opus-4-5}). We set the temperature to the default value of 1.0.

We additionally include a new baseline, \textsc{G$^{2}$Fuzz} \cite{non-textual}, a state-of-the-art coverage-guided greybox fuzzer that augments inputs using LLM-generated test cases. We add this comparison to show that guiding fuzzing with LLM-predicted call stacks can be more effective at triggering the target vulnerability than simply generating inputs with an LLM.
\textsc{AD} \cite{AD} and \textsc{ISC4DGF} \cite{ISC4DGF} are two closely related works; however, the public artifact link for \textsc{AD} is no longer accessible, and \textsc{ISC4DGF} does not provide a public artifact. We contacted the authors of both projects but received no response, and therefore cannot include them in our evaluation.

\subsection{Effectiveness}\label{section-6.2}

\begin{figure}[t!]
    \centering
    \includegraphics[width=\textwidth]{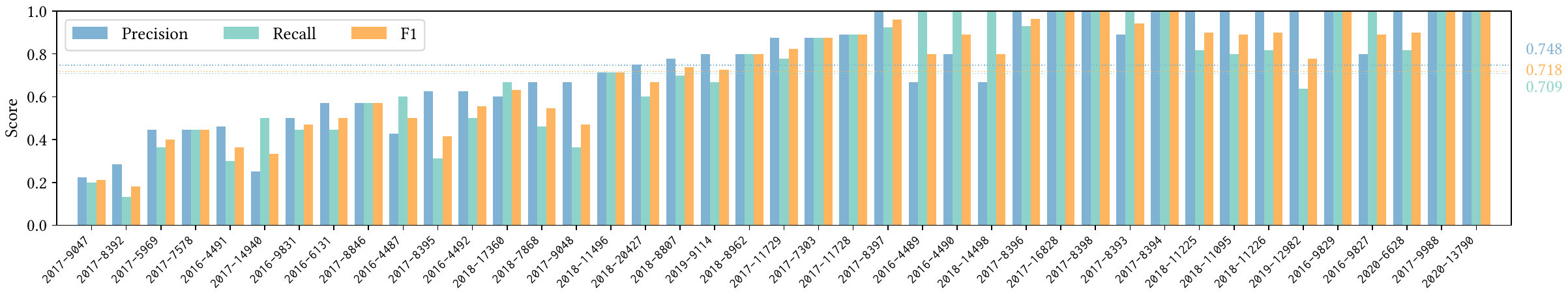}
    \vspace{-0.3in}
    \caption{Accuracy of \textsc{Staczzer}'s call-stack prediction across 41 vulnerabilities. The dotted line represents the average value.}
    \label{acc}
\end{figure}

In \autoref{acc}, we report the accuracy of the call-stack predictions made by the LLM in \textsc{Staczzer}. The results are only slightly lower than those of fully using the code agent shown in \autoref{cfg-acc}, indicating that \textsc{Staczzer} can reduce the time overhead on small programs while still maintaining high accuracy.

\begin{table}[t!]
    \centering
    \caption{Median TTE (including preprocessing) over 40 runs on 41 known vulnerabilities.
\textbf{MTTE} is the median TTE (s); if unavailable, we report $(x/40)$ where $x$ runs trigger the bug.
\textbf{Ratio} is the MTTE relative to \textsc{Staczzer}; if MTTE is unavailable, it is computed using $2\times$ the time limit (48h).
The best fuzzer per vulnerability is in bold.
In \textbf{Performance vs.\ \textsc{Staczzer}}, $a/b$ means \textsc{Staczzer} wins on $a$ vulnerabilities and loses on $b$; this is used for the sign-test p-value shown in the next row.
\textbf{\# Stably reproducible} counts vulnerabilities with a computable MTTE, and \textbf{\# Best performance} counts vulnerabilities where each fuzzer is best.}
    \label{effectiveness}
    \vspace{-0.1in}
    \tiny
    \begin{tabular}{llrrrrrrrrrrr}
        \toprule
        \multicolumn{1}{c}{\multirow{2}{*}{\textbf{Program}}} & \multicolumn{1}{c}{\multirow{2}{*}{\textbf{CVE ID}}} & \multicolumn{1}{c}{\textbf{\textsc{Staczzer}}} & \multicolumn{2}{c}{\textbf{\textsc{AFLGo}}} & \multicolumn{2}{c}{\textbf{\textsc{WindRanger}}} & \multicolumn{2}{c}{\textbf{\textsc{DAFL}}} &  \multicolumn{2}{c}{\textbf{\textsc{G$^2$Fuzz}}} & \multicolumn{2}{c}{\textbf{\textsc{AFL}}} \\ 
        \cmidrule(lr){3-3}\cmidrule(lr){4-5}\cmidrule(lr){6-7}\cmidrule(lr){8-9}\cmidrule(lr){10-11}\cmidrule(lr){12-13}
        & & \multicolumn{1}{c}{\textbf{MTTE}} & \multicolumn{1}{c}{\textbf{MTTE}} & \multicolumn{1}{c}{\textbf{Ratio}} & \multicolumn{1}{c}{\textbf{MTTE}} & \multicolumn{1}{c}{\textbf{Ratio}} & \multicolumn{1}{c}{\textbf{MTTE}} & \multicolumn{1}{c}{\textbf{Ratio}} & \multicolumn{1}{c}{\textbf{MTTE}} & \multicolumn{1}{c}{\textbf{Ratio}} & \multicolumn{1}{c}{\textbf{MTTE}} & \multicolumn{1}{c}{\textbf{Ratio}} \\ 
\midrule\multirow{17}{*}{\texttt{swftophp}} & 2016-9827 & 112 & 461 & 4.12 & \textbf{95} & 0.85 & 147 & 1.31 & 123 & 1.10 & 116 & 1.04 \\
 & 2016-9829 & \textbf{245} & 676 & 2.76 & 501 & 2.04 & 411 & 1.68 & 466 & 1.90 & 429 & 1.75 \\
 & 2016-9831 & 526 & 785 & 1.49 & 459 & 0.87 & \textbf{315} & 0.60 & 607 & 1.15 & 416 & 0.79 \\
 & 2017-9988 & 4,704 & 3,925 & 0.83 & \textbf{3,395} & 0.72 & 4,101 & 0.87 & 4,548 & 0.97 & 3,473 & 0.74 \\
 & 2017-11728 & \textbf{599} & 2,815 & 4.70 & 995 & 1.66 & 631 & 1.05 & 3,231 & 5.39 & 2,364 & 3.95 \\
 & 2017-11729 & \textbf{213} & 978 & 4.59 & 400 & 1.88 & 287 & 1.35 & 1,196 & 5.62 & 826 & 3.88 \\
 & 2017-7578 & 416 & 845 & 2.03 & 680 & 1.63 & 643 & 1.55 & \textbf{333} & 0.80 & 848 & 2.04 \\
 & 2018-7868 & (1/40) & (0/40) & 1.00 & (0/40) & 1.00 & \textbf{(12/40)} & 1.00 & (0/40) & 1.00 & (0/40) & 1.00 \\
 & 2018-8807 & (0/40) & (0/40) & 1.00 & (0/40) & 1.00 & (0/40) & 1.00 & (0/40) & 1.00 & (0/40) & 1.00 \\
 & 2018-8962 & (0/40) & (0/40) & 1.00 & (0/40) & 1.00 & \textbf{(2/40)} & 1.00 & (0/40) & 1.00 & (0/40) & 1.00 \\
 & 2018-11095 & \textbf{837} & 8,339 & 9.96 & 2,949 & 3.52 & 2,699 & 3.22 & 5,542 & 6.62 & 6,873 & 8.21 \\
 & 2018-11225 & \textbf{32,673} & (8/40) & 5.29 & (11/40) & 5.29 & 45,296 & 1.39 & (17/40) & 5.29 & (13/40) & 5.29 \\
 & 2018-11226 & 79,229 & (11/40) & 2.18 & (11/40) & 2.18 & \textbf{62,402} & 0.79 & (10/40) & 2.18 & (11/40) & 2.18 \\
 & 2018-20427 & (18/40) & 11,154 & 0.06 & (11/40) & 1.00 & 24,395 & 0.14 & 14,293 & 0.08 & \textbf{9,846} & 0.06 \\
 & 2019-9114 & \textbf{51,070} & (20/40) & 3.38 & (16/40) & 3.38 & (8/40) & 3.38 & 83,770 & 1.64 & (15/40) & 3.38 \\
 & 2019-12982 & \textbf{40,420} & (14/40) & 4.28 & (13/40) & 4.28 & 71,703 & 1.77 & 51,079 & 1.26 & (18/40) & 4.28 \\
 & 2020-6628 & \textbf{47,681} & (10/40) & 3.62 & (4/40) & 3.62 & (15/40) & 3.62 & (9/40) & 3.62 & (17/40) & 3.62 \\
\midrule\multirow{2}{*}{\texttt{lrzip}} & 2017-8846 & \textbf{(2/40)} & (1/40) & 1.00 & (0/40) & 1.00 & (1/40) & 1.00 & (0/40) & 1.00 & (0/40) & 1.00 \\
 & 2018-11496 & 23 & 193 & 8.39 & \textbf{21} & 0.91 & 25 & 1.09 & \textbf{21} & 0.91 & 35 & 1.52 \\
\midrule\multirow{6}{*}{\texttt{cxxfilt}} & 2016-4487 & 655 & 1,083 & 1.65 & 2,408 & 3.68 & \textbf{574} & 0.88 & 1,000 & 1.53 & 862 & 1.32 \\
 & 2016-4489 & \textbf{502} & 2,628 & 5.24 & 790 & 1.57 & 2,235 & 4.45 & 2,570 & 5.12 & 1,871 & 3.73 \\
 & 2016-4490 & \textbf{205} & 804 & 3.92 & 620 & 3.02 & 1,080 & 5.27 & 594 & 2.90 & 618 & 3.01 \\
 & 2016-4491 & (0/40) & (1/40) & 1.00 & (0/40) & 1.00 & \textbf{(4/40)} & 1.00 & (1/40) & 1.00 & (0/40) & 1.00 \\
 & 2016-4492 & \textbf{3,016} & 3,566 & 1.18 & 5,272 & 1.75 & 3,944 & 1.31 & 6,337 & 2.10 & 3,210 & 1.06 \\
 & 2016-6131 & (0/40) & \textbf{(1/40)} & 1.00 & (0/40) & 1.00 & (0/40) & 1.00 & (0/40) & 1.00 & (0/40) & 1.00 \\
\midrule\multirow{3}{*}{\texttt{objcopy}} & 2017-8393 & \textbf{860} & 4,417 & 5.14 & 3,090 & 3.59 & 1,977 & 2.30 & 3,740 & 4.35 & 3,415 & 3.97 \\
 & 2017-8394 & \textbf{641} & 3,563 & 5.56 & 1,268 & 1.98 & 15,029 & 23.45 & 2,862 & 4.46 & 1,367 & 2.13 \\
 & 2017-8395 & 506 & 1,857 & 3.67 & 221 & 0.44 & 290 & 0.57 & 251 & 0.50 & \textbf{194} & 0.38 \\
\midrule\multirow{5}{*}{\texttt{objdump}} & 2017-8392 & \textbf{326} & 2,209 & 6.78 & 629 & 1.93 & 361 & 1.11 & 439 & 1.35 & 466 & 1.43 \\
 & 2017-8396 & (0/40) & (0/40) & 1.00 & (1/40) & 1.00 & (0/40) & 1.00 & \textbf{(2/40)} & 1.00 & (1/40) & 1.00 \\
 & 2017-8397 & \textbf{77,803} & (17/40) & 2.22 & (18/40) & 2.22 & (7/40) & 2.22 & (10/40) & 2.22 & 84,774 & 1.09 \\
 & 2017-8398 & 8,535 & 6,253 & 0.73 & \textbf{2,897} & 0.34 & 4,834 & 0.57 & 6,936 & 0.81 & 5,822 & 0.68 \\
 & 2018-17360 & (1/40) & (0/40) & 1.00 & (0/40) & 1.00 & (0/40) & 1.00 & (0/40) & 1.00 & \textbf{(4/40)} & 1.00 \\
\midrule\multirow{1}{*}{\texttt{strip}} & 2017-7303 & 663 & 3,805 & 5.74 & 5,708 & 8.61 & 667 & 1.01 & 261 & 0.39 & \textbf{233} & 0.35 \\
\midrule\multirow{1}{*}{\texttt{nm}} & 2017-14940 & (0/40) & (0/40) & 1.00 & (0/40) & 1.00 & (0/40) & 1.00 & (0/40) & 1.00 & (0/40) & 1.00 \\
\midrule\multirow{1}{*}{\texttt{readelf}} & 2017-16828 & \textbf{303} & 1,139 & 3.76 & 816 & 2.69 & 1,723 & 5.69 & 1,148 & 3.79 & 628 & 2.07 \\
\midrule\multirow{3}{*}{\texttt{xmllint}} & 2017-5969 & 1,089 & 3,166 & 2.91 & 4,647 & 4.27 & 1,410 & 1.29 & 683 & 0.63 & \textbf{665} & 0.61 \\
 & 2017-9047 & \textbf{69,109} & (5/40) & 2.50 & (2/40) & 2.50 & (13/40) & 2.50 & (4/40) & 2.50 & (5/40) & 2.50 \\
 & 2017-9048 & 18,843 & (1/40) & 9.17 & (1/40) & 9.17 & \textbf{13,093} & 0.69 & (5/40) & 9.17 & (4/40) & 9.17 \\
\midrule\multirow{2}{*}{\texttt{cjpeg}} & 2018-14498 & (8/40) & (2/40) & 1.00 & (0/40) & 1.00 & \textbf{23,156} & 0.13 & (14/40) & 1.00 & (11/40) & 1.00 \\
 & 2020-13790 & \textbf{(6/40)} & (1/40) & 1.00 & (4/40) & 1.00 & \textbf{(6/40)} & 1.00 & (1/40) & 1.00 & (2/40) & 1.00 \\

    \midrule
         \multicolumn{2}{c}{\textbf{Average speedup}} & \multicolumn{1}{c}{-} & \multicolumn{2}{c}{3.14} & \multicolumn{2}{c}{2.26} & \multicolumn{2}{c}{2.13} & \multicolumn{2}{c}{2.23} & \multicolumn{2}{c}{2.13} \\
         \multicolumn{2}{c}{\textbf{Performance vs. \textsc{Staczzer}}} & \multicolumn{1}{c}{-} & \multicolumn{2}{c}{\textbf{32}/5} & \multicolumn{2}{c}{\textbf{29}/7} & \multicolumn{2}{c}{\textbf{24}/12} & \multicolumn{2}{c}{\textbf{26}/11} & \multicolumn{2}{c}{\textbf{26}/10} \\
         \multicolumn{2}{c}{\textbf{P-value in the sign test}} & \multicolumn{1}{c}{-} & \multicolumn{2}{c}{$4 \times 10^{-6}$} & \multicolumn{2}{c}{$2 \times 10^{-4}$} & \multicolumn{2}{c}{0.03} & \multicolumn{2}{c}{0.01} & \multicolumn{2}{c}{0.006} \\
         \multicolumn{2}{c}{\textbf{\# Stably reproducible}} & \multicolumn{1}{c}{\textbf{29}} & \multicolumn{2}{c}{22} & \multicolumn{2}{c}{21} & \multicolumn{2}{c}{27} & \multicolumn{2}{c}{24} & \multicolumn{2}{c}{23} \\
         \multicolumn{2}{c}{\textbf{\# Best performance}} & \multicolumn{1}{c}{\textbf{19}} & \multicolumn{2}{c}{1} & \multicolumn{2}{c}{4} & \multicolumn{2}{c}{9} & \multicolumn{2}{c}{3} & \multicolumn{2}{c}{5} \\
         \bottomrule
    \end{tabular}
\end{table}

In \autoref{effectiveness}, we present the complete median TTE results of \textsc{Staczzer} and the baselines on 41 vulnerabilities. \textsc{Staczzer} demonstrates average speedups of $2.13\times$ to $3.14\times$ compared to the baselines in vulnerability reproduction, along with win-loss ratios of $2\times$ to $6.4\times$, fully demonstrating the effectiveness of \textsc{Staczzer}. The p-values from the one-tailed sign test based on win-loss counts are all below 0.05, indicating that \textsc{Staczzer} significantly outperforms all baselines. \textsc{Staczzer} can stably reproduce 29 vulnerabilities, outperforming all other baselines. Moreover, the \textbf{\# Best performance} row shows that \textsc{Staczzer} achieves the best performance on 19 vulnerabilities, which is $2.11\times$ to $19\times$ that of other baselines. 

\begin{figure}[t!]
    \centering
    \includegraphics[width=\textwidth]{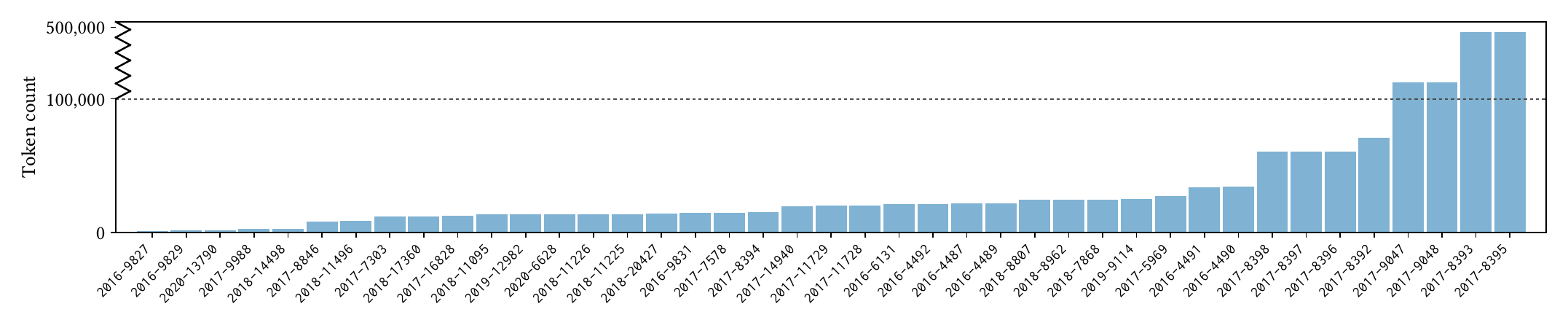}
    \vspace{-0.35in}
    \caption{Token counts of function slices in the call graph that can reach the target for the 41 vulnerabilities.}
    \label{token}
\end{figure}

We further show in \autoref{token} the token counts of call-graph-reachable function slices in \textsc{Staczzer}. For most vulnerabilities, slices stay within our $10^5$-token limit; only four exceed it. In \autoref{agent}, we report the code-agent token and time overhead for these four cases. Overall, \textsc{Staczzer}'s LLM token and time costs are acceptable.

\textbf{\uline{Overall, \textsc{Staczzer} attains high accuracy in call-stack prediction, thereby providing strong acceleration for DGF.}}

\begin{table}[t!]
\scriptsize
\centering
\caption{Token count and time overhead of the code agent prediction on the remaining 4 vulnerabilities.}
\label{agent}
\vspace{-0.1in}
\begin{tabular}{lrrrr}
\toprule
\multicolumn{1}{c}{\textbf{Metric}} &
\multicolumn{1}{c}{\textbf{\texttt{2017-8393}}} & \multicolumn{1}{c}{\textbf{\texttt{2017-8395}}} & \multicolumn{1}{c}{\textbf{\texttt{2017-9047}}} & \multicolumn{1}{c}{\textbf{\texttt{2017-9048}}} \\
\midrule
\textbf{Token count} & 1,462,307 & 999,598 & 2,297,026 & 1,611,668 \\
\textbf{Time overhead (s)} & 260 & 233 & 303 & 177 \\
\bottomrule
\end{tabular}
\end{table}

\subsection{Impact of Different LLMs}\label{section-6.3}

\begin{figure}[t!]
    \centering
    \includegraphics[width=\textwidth]{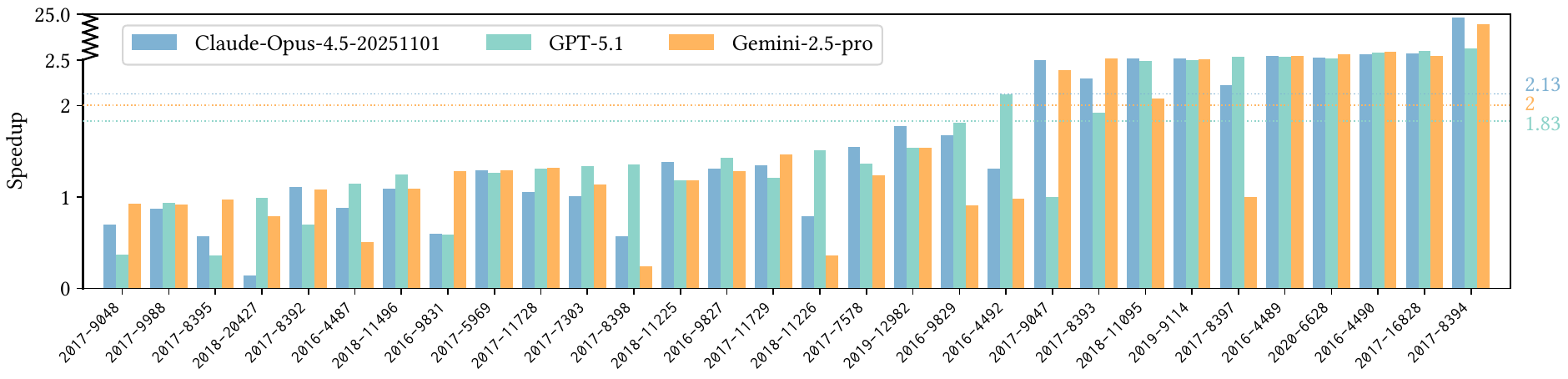}
    \vspace{-0.3in}
    \caption{Speedup over DAFL under different LLM settings. The dotted line represents the average value. For brevity, vulnerabilities that no fuzzer can stably reproduce are omitted.}
    \label{llm}
\end{figure}

We additionally evaluated two representative LLMs: (1) GPT-5.1 \cite{gpt-5.1} for \textsc{SingleTurnPrediction} and Codex~0.75.0 \cite{codex} (GPT-5.1-codex-max high \cite{gpt-5.1-codex}) for \textsc{CodeAgentPrediction}; (2) Gemini-2.5-pro \cite{gemini-2.5-pro} for \textsc{SingleTurnPrediction} and Gemini CLI~0.25.2 \cite{gemini-cli} (Gemini-2.5-pro \cite{gemini-2.5-pro}) for \textsc{CodeAgentPrediction}. We set the temperature to its default value of 1.0. GPT-5.1 and Gemini-2.5-pro achieve average (Precision, Recall, F1) scores of
(0.750, 0.694, 0.714) and (0.716, 0.705, 0.701), respectively. \autoref{llm} reports the speedup of \textsc{Staczzer} over the state-of-the-art directed greybox fuzzer DAFL when instantiated with three different LLMs. The average speedup ranges from $1.83\times$ to $2.13\times$; all three variants substantially outperform the baseline and exhibit no significant differences. \textbf{\uline{These results indicate that \textsc{Staczzer} is not tied to a specific LLM, but achieves strong performance across diverse LLM backends.}}

\subsection{Ablation Study}\label{section-6.4}

\begin{figure}[t!]
    \centering
    \includegraphics[width=\textwidth]{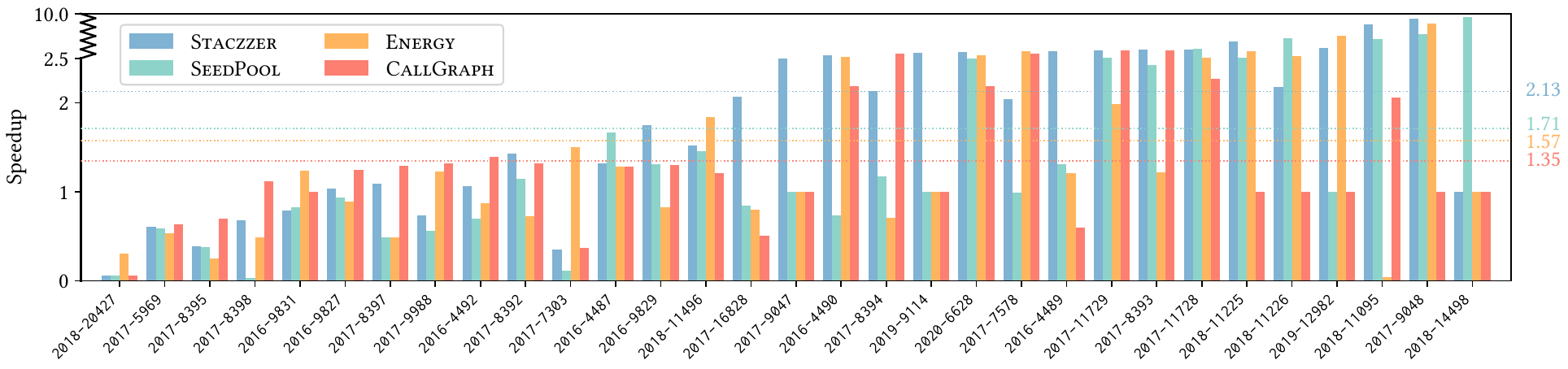}
    \vspace{-0.3in}
    \caption{Speedup over AFL under different ablation settings. The dotted line represents the average value. For brevity, vulnerabilities that no fuzzer can stably reproduce are omitted.}
    \label{ablation}
\end{figure}

We conduct an ablation study on the components of \textsc{Staczzer}: (1) \textsc{SeedPool} denotes enabling only seed selection, i.e., we do not modify \lineref{tdgf:assign} and instead use AFL's original power-scheduling strategy; (2) \textsc{Energy} denotes enabling only power scheduling, i.e., we do not modify \lineref{tdgf:choose} and instead use AFL's original seed-selection strategy; and (3) \textsc{CallGraph} denotes using all call sites in the call graph that can potentially reach the target as the guidance trace, rather than the LLM-predicted call stack. \autoref{ablation} reports the speedup of \textsc{Staczzer} and these three configurations over the foundation fuzzer AFL. For \textsc{SeedPool} and \textsc{Energy}, both yield clear speedups, but neither matches the improvement achieved by \textsc{Staczzer} when both are enabled, indicating that each modification contributes to the overall effectiveness. For \textsc{CallGraph}, the speedup is only marginal, suggesting that call-graph-based trace guidance alone provides imprecise signals, and that refining it into a call stack via the LLM is key to the performance gains. \textbf{\uline{Overall, seed selection, power scheduling, and LLM-based call-stack prediction are all effective and necessary components.}}

\subsection{New Bugs}\label{section-6.5}

\begin{table}[t!]
    \centering
    \caption{New bugs found through regression patch testing on the latest versions of benchmarks from the controlled experiments.}
    \label{new-bugs}
    \vspace{-0.1in}
    \scriptsize
    \begin{tabular}{lccc}
        \toprule
        \textbf{Program} & \textbf{Location} & \textbf{CWE} & \textbf{CVE} \\ 
        \midrule\multirow{2}{*}{\texttt{objdump}} & \texttt{objdump.c:4501} & Out-of-bounds Read & CVE-2025-11081 \\ 
        ~ & \texttt{prdbg.c:2452} & Unexpected Status Code or Return Value & CVE-2025-11839 \\ 
        \midrule\multirow{8}{*}{\texttt{ld}} & \texttt{elf-eh-frame.c:756} & Heap-based Buffer Overflow & CVE-2025-11082 \\ 
        ~ & \texttt{cache.c:435} & Heap-based Buffer Overflow & CVE-2025-11083 \\ 
        ~ & \texttt{elflink.c:14898}  & Out-of-bounds Read & CVE-2025-11412 \\ 
        ~ & \texttt{libbfd.c:989} & Heap-based Buffer Overflow & CVE-2025-11413 \\ 
        ~ & \texttt{elflink.c:115} & Out-of-bounds Read & CVE-2025-11414 \\ 
        ~ & \texttt{elf-eh-frame.c:2170} & Out-of-bounds Read & CVE-2025-11494 \\ 
        ~ & \texttt{elf64-x86-64.c:4507} & Heap-based Buffer Overflow & CVE-2025-11495 \\ 
        ~ & \texttt{ldmisc.c:527} & Out-of-bounds Read & CVE-2025-11840 \\ \bottomrule
    \end{tabular}
\end{table}

\begin{table}[t!]
    \centering
    \caption{Incomplete fixes discovered through incremental patch testing on the patched versions of newly found vulnerabilities in \autoref{new-bugs}.}
    \label{incomplete-fixes}
    \vspace{-0.1in}
    \scriptsize
    \begin{tabular}{lccc}
        \toprule
        \textbf{Program} & \textbf{Location} & \textbf{CWE} & \textbf{Original CVE} \\ 
        \midrule\multirow{2}{*}{\texttt{ld}} & \texttt{elf-eh-frame.c:766 } & Heap-based Buffer Overflow & CVE-2025-11082 \\ 
        ~ & \texttt{libbfd.c:989} & Out-of-bounds Write & CVE-2025-11413 \\ 
        \bottomrule
    \end{tabular}
\end{table}

We use \textsc{Staczzer} to regression-test the latest project versions at patch locations for known fixed vulnerabilities from the controlled experiments in \autoref{benchmark}, with a 24-hour time limit. As shown in \autoref{new-bugs}, we discover 10 new vulnerabilities in Binutils~2.45, located in \texttt{objdump} and the GNU linker \texttt{ld}. We reported all of them to the developers; they have been fixed and assigned CVE IDs. For these new vulnerabilities, we further perform incremental patch testing on the patched programs and identify two additional incomplete fixes, as shown in \autoref{incomplete-fixes}; the developers confirmed and fixed them. We list the CWE (Common Weakness Enumeration) for each vulnerability in both tables. These issues in widely deployed software can have severe impact: heap-based buffer overflows and out-of-bounds writes enable illegal memory writes that can disrupt execution or even allow privilege escalation, while out-of-bounds reads can enable illegal memory access and information leakage. \textbf{\uline{These findings demonstrate \textsc{Staczzer}'s practical value in real-world settings.}}


\section{Threats to Validity}\label{section-7}

We discuss potential threats to validity in our experiments.

The main threat to internal validity is fuzzing randomness. Prior work~\cite{evaluatingDGF} shows that at least 40 runs are needed to reduce DGF randomness to an acceptable level. We therefore run each fuzzer on each target 40 times, for a total cost of 58.4 CPU-years, which we believe largely mitigates randomness.

The main threats to external validity are that our approach ignores the order and frequency of executed call sites and does not handle vulnerabilities reachable through multiple call stacks, which can lead to suboptimal seed prioritization. Capturing full execution sequences or enumerating call stacks would significantly increase fuzzer overhead, so our design makes an explicit efficiency-precision trade-off. \textsc{Staczzer} is effective on the benchmark suite used by the SOTA DGF tool \textsc{DAFL} and also finds previously unknown vulnerabilities in real-world critical software that were later fixed, suggesting that its effectiveness generalizes to practical DGF settings.
\section{Related Work}

Our approach is related to (1) directed greybox fuzzing and (2) techniques that leverage LLMs to enhance fuzzing. We summarize the related work below.

\textit{Directed greybox fuzzing.} \textsc{AFLGo}~\cite{AFLGo} pioneered DGF by prioritizing seeds using control-flow-graph distances to targets. Follow-up systems refine this idea from different angles, e.g., value-gap awareness (\textsc{CAFL}~\cite{CAFL}), deviation blocks (\textsc{WindRanger}~\cite{Windranger}), Monte Carlo guidance (\textsc{MC$^2$}~\cite{MC2}), selective instrumentation (\textsc{SelectFuzz}~\cite{SelectFuzz}), value-flow guidance (\textsc{DAFL}~\cite{DAFL}), dynamic indirect-call resolution (\textsc{PDGF}~\cite{PDGF}), combined control/data-flow criticality (\textsc{WAFLGo}~\cite{WAFLGo}), and learning-based mutation (\textsc{IDFuzz}~\cite{IDFuzz}). All these methods depend on imprecise static-analysis distance metrics. \textsc{Hawkeye} \cite{Hawkeye} and \textsc{SDFuzz} \cite{SDFuzz} incorporate target call stacks, but call stacks are hard to obtain and they still rely on imprecise distance metrics. Our experiments show \textsc{Staczzer}'s predicted call stack is a stronger metric. \textsc{Halo}~\cite{Halo} and \textsc{DeepGo}~\cite{DeepGo} mitigate distance imprecision via invariant inference and reinforcement learning, respectively, but they are not open-source, preventing us from comparing with them. \textsc{Beacon}~\cite{Beacon} prunes paths proven irrelevant by static analysis, complementing \textsc{Staczzer}'s seed prioritization. Multi-target DGF includes alarm-driven fuzzing~\cite{Prospector, FishFuzz, ParmeSan, SAVIOR} and critical-set fuzzing~\cite{Titan, LeoFuzz, AFLRun, Lyso}; both rely on conventional distance metrics, and can incorporate \textsc{Staczzer}'s call-stack metric.

\textit{Fuzzing with LLMs.} LLMs are widely used to generate initial corpora and mutators for structured-input fuzzing (e.g., compilers/interpreters/theorem provers)~\cite{Fuzz4All, llm-solver, fuzzing-js}, non-textual inputs~\cite{non-textual}, deep-learning libraries~\cite{TitanFuzz}, and protocols~\cite{ChatAFL}, and to synthesize harnesses~\cite{PromptFuzz}. They also assist directed fuzzing by producing vulnerability-triggering seeds~\cite{Magneto, ISC4DGF}. \textsc{Staczzer} is orthogonal and can be combined with these approaches.
AD \cite{AD} uses LLMs to rescale static-analysis distances in DGF, but the metric remains imprecise due to inherent over-approximation; \textsc{Staczzer} instead uses predicted guidance traces to mitigate this issue.

\section{Conclusion}
To address the imprecision of static-analysis-based distance metrics in directed greybox fuzzing (DGF), we propose trace-guided directed greybox fuzzing (TDGF), which prioritizes seeds whose execution traces overlap more with a guidance trace. We study two guidance-trace types and find that coarse-grained call-stack traces provide guidance comparable to fine-grained control-flow traces while being easier for large language models (LLMs) to predict. Based on this finding, we present \textsc{Staczzer}, which uses LLM-predicted vulnerability-triggering call stacks to guide seed prioritization in DGF. We evaluate \textsc{Staczzer} with experiments totaling 58.4 CPU-years: it reproduces known vulnerabilities with average speedups of $2.13\times$ to $3.14\times$ over baselines, and it discovers 10 new vulnerabilities and 2 incomplete fixes, with 10 assigned CVE IDs. These results show that \textsc{Staczzer} substantially improves directed vulnerability triggering over existing directed fuzzers.

\bibliographystyle{ACM-Reference-Format}
\bibliography{sample-base}

@String{Computer = "{IEEE} Computer" }

@inproceedings{FishFuzz,
  author       = {Han Zheng and
                  Jiayuan Zhang and
                  Yuhang Huang and
                  Zezhong Ren and
                  He Wang and
                  Chunjie Cao and
                  Yuqing Zhang and
                  Flavio Toffalini and
                  Mathias Payer},
  editor       = {Joseph A. Calandrino and
                  Carmela Troncoso},
  title        = {{FISHFUZZ:} Catch Deeper Bugs by Throwing Larger Nets},
  booktitle    = {32nd {USENIX} Security Symposium, {USENIX} Security 2023, Anaheim,
                  CA, USA, August 9-11, 2023},
  pages        = {1343--1360},
  publisher    = {{USENIX} Association},
  year         = {2023},
  url          = {https://www.usenix.org/conference/usenixsecurity23/presentation/zheng},
  bibsource    = {dblp computer science bibliography, https://dblp.org}
}

@inproceedings{Prospector,
  author       = {Zhijie Zhang and
                  Liwei Chen and
                  Haolai Wei and
                  Gang Shi and
                  Dan Meng},
  editor       = {Maria Christakis and
                  Michael Pradel},
  title        = {Prospector: Boosting Directed Greybox Fuzzing for Large-Scale Target
                  Sets with Iterative Prioritization},
  booktitle    = {Proceedings of the 33rd {ACM} {SIGSOFT} International Symposium on
                  Software Testing and Analysis, {ISSTA} 2024, Vienna, Austria, September
                  16-20, 2024},
  pages        = {1351--1363},
  publisher    = {{ACM}},
  year         = {2024},
  url          = {https://doi.org/10.1145/3650212.3680365},
  doi          = {10.1145/3650212.3680365},
  bibsource    = {dblp computer science bibliography, https://dblp.org}
}

@inproceedings{ParmeSan,
  author       = {Sebastian {\"{O}}sterlund and
                  Kaveh Razavi and
                  Herbert Bos and
                  Cristiano Giuffrida},
  editor       = {Srdjan Capkun and
                  Franziska Roesner},
  title        = {ParmeSan: Sanitizer-guided Greybox Fuzzing},
  booktitle    = {29th {USENIX} Security Symposium, {USENIX} Security 2020, August 12-14,
                  2020},
  pages        = {2289--2306},
  publisher    = {{USENIX} Association},
  year         = {2020},
  url          = {https://www.usenix.org/conference/usenixsecurity20/presentation/osterlund},
  bibsource    = {dblp computer science bibliography, https://dblp.org}
}

@inproceedings{SAVIOR,
  author       = {Yaohui Chen and
                  Peng Li and
                  Jun Xu and
                  Shengjian Guo and
                  Rundong Zhou and
                  Yulong Zhang and
                  Tao Wei and
                  Long Lu},
  title        = {{SAVIOR:} Towards Bug-Driven Hybrid Testing},
  booktitle    = {2020 {IEEE} Symposium on Security and Privacy, {SP} 2020, San Francisco,
                  CA, USA, May 18-21, 2020},
  pages        = {1580--1596},
  publisher    = {{IEEE}},
  year         = {2020},
  url          = {https://doi.org/10.1109/SP40000.2020.00002},
  doi          = {10.1109/SP40000.2020.00002},
  bibsource    = {dblp computer science bibliography, https://dblp.org}
}

@inproceedings{ASan,
  author       = {Konstantin Serebryany and
                  Derek Bruening and
                  Alexander Potapenko and
                  Dmitriy Vyukov},
  editor       = {Gernot Heiser and
                  Wilson C. Hsieh},
  title        = {AddressSanitizer: {A} Fast Address Sanity Checker},
  booktitle    = {Proceedings of the 2012 {USENIX} Annual Technical Conference, {USENIX}
                  {ATC} 2012, Boston, MA, USA, June 13-15, 2012},
  pages        = {309--318},
  publisher    = {{USENIX} Association},
  year         = {2012},
  url          = {https://www.usenix.org/conference/atc12/technical-sessions/presentation/serebryany},
  bibsource    = {dblp computer science bibliography, https://dblp.org}
}

@inproceedings{AFLGo,
  author       = {Marcel B{\"{o}}hme and
                  Van{-}Thuan Pham and
                  Manh{-}Dung Nguyen and
                  Abhik Roychoudhury},
  editor       = {Bhavani Thuraisingham and
                  David Evans and
                  Tal Malkin and
                  Dongyan Xu},
  title        = {Directed Greybox Fuzzing},
  booktitle    = {Proceedings of the 2017 {ACM} {SIGSAC} Conference on Computer and
                  Communications Security, {CCS} 2017, Dallas, TX, USA, October 30 -
                  November 03, 2017},
  pages        = {2329--2344},
  publisher    = {{ACM}},
  year         = {2017},
  url          = {https://doi.org/10.1145/3133956.3134020},
  doi          = {10.1145/3133956.3134020},
  bibsource    = {dblp computer science bibliography, https://dblp.org}
}

@misc{AFL,
  title        = {American Fuzzy Lop},
  author       = {Micha{\l} Zalewski},
  year         = {2013},
  howpublished = {\url{https://lcamtuf.coredump.cx/afl/}}
}

@misc{CVE-2016-4489,
  author       = {MITRE},
  title        = {CVE-2016-4489},
  year         = {2016},
  howpublished = {\url{https://cve.mitre.org/cgi-bin/cvename.cgi?name=CVE-2016-4489}}
}

@misc{claude-opus-4-5,
  title = {Claude-Opus-4.5},
  author = {{Anthropic}},
  year = {2025},
  howpublished = {\url{https://www.anthropic.com/news/claude-opus-4-5}},
}

@misc{claude-code,
  title = {Claude Code},
  author = {{Anthropic}},
  year = {2025},
  howpublished = {\url{https://claude.com/product/claude-code}},
}

@misc{codex,
  title = {Codex},
  author = {{OpenAI}},
  year = {2025},
  howpublished = {\url{https://openai.com/codex/}},
}

@misc{gpt-5.1,
  title = {GPT-5.1},
  author = {{OpenAI}},
  year = {2025},
  howpublished = {\url{https://openai.com/index/gpt-5-1/}},
}

@misc{gpt-5.1-codex,
  title = {GPT-5.1-Codex-Max},
  author = {{OpenAI}},
  year = {2025},
  howpublished = {\url{https://openai.com/index/gpt-5-1-codex-max/}},
}

@misc{gemini-cli,
  title = {Gemini CLI},
  author = {{Google}},
  year = {2025},
  howpublished = {\url{https://geminicli.com}},
}

@misc{gemini-2.5-pro,
  title = {Gemini-2.5-pro},
  author = {{Google}},
  year = {2025},
  howpublished = {\url{https://docs.cloud.google.com/vertex-ai/generative-ai/docs/models/gemini/2-5-pro}},
}

@inproceedings{SVF,
  author       = {Yulei Sui and
                  Jingling Xue},
  editor       = {Ayal Zaks and
                  Manuel V. Hermenegildo},
  title        = {{SVF:} interprocedural static value-flow analysis in {LLVM}},
  booktitle    = {Proceedings of the 25th International Conference on Compiler Construction,
                  {CC} 2016, Barcelona, Spain, March 12-18, 2016},
  pages        = {265--266},
  publisher    = {{ACM}},
  year         = {2016},
  url          = {https://doi.org/10.1145/2892208.2892235},
  doi          = {10.1145/2892208.2892235},
  bibsource    = {dblp computer science bibliography, https://dblp.org}
}

@article{LeoFuzz,
  author       = {Hongliang Liang and
                  Xinglin Yu and
                  Xianglin Cheng and
                  Jie Liu and
                  Jin Li},
  title        = {Multiple Targets Directed Greybox Fuzzing},
  journal      = {{IEEE} Trans. Dependable Secur. Comput.},
  volume       = {21},
  number       = {1},
  pages        = {325--339},
  year         = {2024},
  url          = {https://doi.org/10.1109/TDSC.2023.3253120},
  doi          = {10.1109/TDSC.2023.3253120},
  bibsource    = {dblp computer science bibliography, https://dblp.org}
}

@inproceedings{Titan,
  author       = {Heqing Huang and
                  Peisen Yao and
                  Hung{-}Chun Chiu and
                  Yiyuan Guo and
                  Charles Zhang},
  title        = {Titan : Efficient Multi-target Directed Greybox Fuzzing},
  booktitle    = {{IEEE} Symposium on Security and Privacy, {SP} 2024, San Francisco,
                  CA, USA, May 19-23, 2024},
  pages        = {1849--1864},
  publisher    = {{IEEE}},
  year         = {2024},
  url          = {https://doi.org/10.1109/SP54263.2024.00059},
  doi          = {10.1109/SP54263.2024.00059},
  bibsource    = {dblp computer science bibliography, https://dblp.org}
}

@inproceedings{AFLRun,
  author       = {Huanyao Rong and
                  Wei You and
                  Xiaofeng Wang and
                  Tianhao Mao},
  editor       = {Davide Balzarotti and
                  Wenyuan Xu},
  title        = {Toward Unbiased Multiple-Target Fuzzing with Path Diversity},
  booktitle    = {33rd {USENIX} Security Symposium, {USENIX} Security 2024, Philadelphia,
                  PA, USA, August 14-16, 2024},
  publisher    = {{USENIX} Association},
  year         = {2024},
  url          = {https://www.usenix.org/conference/usenixsecurity24/presentation/rong},
  bibsource    = {dblp computer science bibliography, https://dblp.org}
}

@inproceedings{Windranger,
  author       = {Zhengjie Du and
                  Yuekang Li and
                  Yang Liu and
                  Bing Mao},
  title        = {Windranger: {A} Directed Greybox Fuzzer driven by Deviation Basic
                  Blocks},
  booktitle    = {44th {IEEE/ACM} 44th International Conference on Software Engineering,
                  {ICSE} 2022, Pittsburgh, PA, USA, May 25-27, 2022},
  pages        = {2440--2451},
  publisher    = {{ACM}},
  year         = {2022},
  url          = {https://doi.org/10.1145/3510003.3510197},
  doi          = {10.1145/3510003.3510197},
  bibsource    = {dblp computer science bibliography, https://dblp.org}
}

@inproceedings{Hawkeye,
  author       = {Hongxu Chen and
                  Yinxing Xue and
                  Yuekang Li and
                  Bihuan Chen and
                  Xiaofei Xie and
                  Xiuheng Wu and
                  Yang Liu},
  editor       = {David Lie and
                  Mohammad Mannan and
                  Michael Backes and
                  XiaoFeng Wang},
  title        = {Hawkeye: Towards a Desired Directed Grey-box Fuzzer},
  booktitle    = {Proceedings of the 2018 {ACM} {SIGSAC} Conference on Computer and
                  Communications Security, {CCS} 2018, Toronto, ON, Canada, October
                  15-19, 2018},
  pages        = {2095--2108},
  publisher    = {{ACM}},
  year         = {2018},
  url          = {https://doi.org/10.1145/3243734.3243849},
  doi          = {10.1145/3243734.3243849},
  bibsource    = {dblp computer science bibliography, https://dblp.org}
}

@inproceedings{SelectFuzz,
  author       = {Changhua Luo and
                  Wei Meng and
                  Penghui Li},
  title        = {SelectFuzz: Efficient Directed Fuzzing with Selective Path Exploration},
  booktitle    = {44th {IEEE} Symposium on Security and Privacy, {SP} 2023, San Francisco,
                  CA, USA, May 21-25, 2023},
  pages        = {2693--2707},
  publisher    = {{IEEE}},
  year         = {2023},
  url          = {https://doi.org/10.1109/SP46215.2023.10179296},
  doi          = {10.1109/SP46215.2023.10179296},
  bibsource    = {dblp computer science bibliography, https://dblp.org}
}

@inproceedings{SDFuzz,
  author       = {Penghui Li and
                  Wei Meng and
                  Chao Zhang},
  editor       = {Davide Balzarotti and
                  Wenyuan Xu},
  title        = {SDFuzz: Target States Driven Directed Fuzzing},
  booktitle    = {33rd {USENIX} Security Symposium, {USENIX} Security 2024, Philadelphia,
                  PA, USA, August 14-16, 2024},
  publisher    = {{USENIX} Association},
  year         = {2024},
  url          = {https://www.usenix.org/conference/usenixsecurity24/presentation/li-penghui},
  bibsource    = {dblp computer science bibliography, https://dblp.org}
}

@inproceedings{CAFL,
  author       = {Gwangmu Lee and
                  Woochul Shim and
                  Byoungyoung Lee},
  editor       = {Michael D. Bailey and
                  Rachel Greenstadt},
  title        = {Constraint-guided Directed Greybox Fuzzing},
  booktitle    = {30th {USENIX} Security Symposium, {USENIX} Security 2021, August 11-13,
                  2021},
  pages        = {3559--3576},
  publisher    = {{USENIX} Association},
  year         = {2021},
  url          = {https://www.usenix.org/conference/usenixsecurity21/presentation/lee-gwangmu},
  bibsource    = {dblp computer science bibliography, https://dblp.org}
}

@inproceedings{DAFL,
  author       = {Tae Eun Kim and
                  Jaeseung Choi and
                  Kihong Heo and
                  Sang Kil Cha},
  editor       = {Joseph A. Calandrino and
                  Carmela Troncoso},
  title        = {{DAFL:} Directed Grey-box Fuzzing guided by Data Dependency},
  booktitle    = {32nd {USENIX} Security Symposium, {USENIX} Security 2023, Anaheim,
                  CA, USA, August 9-11, 2023},
  pages        = {4931--4948},
  publisher    = {{USENIX} Association},
  year         = {2023},
  url          = {https://www.usenix.org/conference/usenixsecurity23/presentation/kim-tae-eun},
  bibsource    = {dblp computer science bibliography, https://dblp.org}
}

@inproceedings{Beacon,
  author       = {Heqing Huang and
                  Yiyuan Guo and
                  Qingkai Shi and
                  Peisen Yao and
                  Rongxin Wu and
                  Charles Zhang},
  title        = {{BEACON:} Directed Grey-Box Fuzzing with Provable Path Pruning},
  booktitle    = {43rd {IEEE} Symposium on Security and Privacy, {SP} 2022, San Francisco,
                  CA, USA, May 22-26, 2022},
  pages        = {36--50},
  publisher    = {{IEEE}},
  year         = {2022},
  url          = {https://doi.org/10.1109/SP46214.2022.9833751},
  doi          = {10.1109/SP46214.2022.9833751},
  bibsource    = {dblp computer science bibliography, https://dblp.org}
}

@book{SignTest,
  title={Nonparametric statistical methods},
  author={Hollander, Myles and Wolfe, Douglas A and Chicken, Eric},
  year={2013},
  publisher={John Wiley \& Sons}
}

@article{Rice,
  title={Classes of recursively enumerable sets and their decision problems},
  author={Rice, Henry Gordon},
  journal={Transactions of the American Mathematical society},
  volume={74},
  number={2},
  pages={358--366},
  year={1953},
  publisher={JSTOR}
}

@inproceedings{MC2,
  author       = {Abhishek Shah and
                  Dongdong She and
                  Samanway Sadhu and
                  Krish Singal and
                  Peter Coffman and
                  Suman Jana},
  editor       = {Heng Yin and
                  Angelos Stavrou and
                  Cas Cremers and
                  Elaine Shi},
  title        = {{MC2:} Rigorous and Efficient Directed Greybox Fuzzing},
  booktitle    = {Proceedings of the 2022 {ACM} {SIGSAC} Conference on Computer and
                  Communications Security, {CCS} 2022, Los Angeles, CA, USA, November
                  7-11, 2022},
  pages        = {2595--2609},
  publisher    = {{ACM}},
  year         = {2022},
  url          = {https://doi.org/10.1145/3548606.3560648},
  doi          = {10.1145/3548606.3560648},
  bibsource    = {dblp computer science bibliography, https://dblp.org}
}

@inproceedings{PDGF,
  author       = {Yujian Zhang and
                  Yaokun Liu and
                  Jinyu Xu and
                  Yanhao Wang},
  title        = {Predecessor-aware Directed Greybox Fuzzing},
  booktitle    = {{IEEE} Symposium on Security and Privacy, {SP} 2024, San Francisco,
                  CA, USA, May 19-23, 2024},
  pages        = {1884--1900},
  publisher    = {{IEEE}},
  year         = {2024},
  url          = {https://doi.org/10.1109/SP54263.2024.00040},
  doi          = {10.1109/SP54263.2024.00040},
  bibsource    = {dblp computer science bibliography, https://dblp.org}
}

@inproceedings{Halo,
  author       = {Heqing Huang and
                  Anshunkang Zhou and
                  Mathias Payer and
                  Charles Zhang},
  title        = {Everything is Good for Something: Counterexample-Guided Directed Fuzzing
                  via Likely Invariant Inference},
  booktitle    = {{IEEE} Symposium on Security and Privacy, {SP} 2024, San Francisco,
                  CA, USA, May 19-23, 2024},
  pages        = {1956--1973},
  publisher    = {{IEEE}},
  year         = {2024},
  url          = {https://doi.org/10.1109/SP54263.2024.00142},
  doi          = {10.1109/SP54263.2024.00142},
  bibsource    = {dblp computer science bibliography, https://dblp.org}
}

@inproceedings{DeepGo,
  author       = {Peihong Lin and
                  Pengfei Wang and
                  Xu Zhou and
                  Wei Xie and
                  Gen Zhang and
                  Kai Lu},
  title        = {DeepGo: Predictive Directed Greybox Fuzzing},
  booktitle    = {31st Annual Network and Distributed System Security Symposium, {NDSS}
                  2024, San Diego, California, USA, February 26 - March 1, 2024},
  publisher    = {The Internet Society},
  year         = {2024},
  url          = {https://www.ndss-symposium.org/ndss-paper/deepgo-predictive-directed-greybox-fuzzing/},
  bibsource    = {dblp computer science bibliography, https://dblp.org}
}

@inproceedings{WAFLGo,
  author       = {Yi Xiang and
                  Xuhong Zhang and
                  Peiyu Liu and
                  Shouling Ji and
                  Xiao Xiao and
                  Hong Liang and
                  Jiacheng Xu and
                  Wenhai Wang},
  editor       = {Davide Balzarotti and
                  Wenyuan Xu},
  title        = {Critical Code Guided Directed Greybox Fuzzing for Commits},
  booktitle    = {33rd {USENIX} Security Symposium, {USENIX} Security 2024, Philadelphia,
                  PA, USA, August 14-16, 2024},
  publisher    = {{USENIX} Association},
  year         = {2024},
  url          = {https://www.usenix.org/conference/usenixsecurity24/presentation/xiang-yi},
  bibsource    = {dblp computer science bibliography, https://dblp.org}
}

@inproceedings{IDFuzz,
  title={$\{$IDFuzz$\}$: Intelligent Directed Grey-box Fuzzing},
  author={Chen, Yiyang and Zhang, Chao and Wang, Long and Zhu, Wenyu and Luo, Changhua and Gui, Nuoqi and Ma, Zheyu and Zhang, Xingjian and Su, Bingkai},
  booktitle={34th USENIX Security Symposium (USENIX Security 25)},
  pages={6219--6238},
  year={2025},
  url          = {https://www.usenix.org/system/files/usenixsecurity25-chen-yiyang.pdf},
}

@inproceedings{Fuzz4All,
  author       = {Chunqiu Steven Xia and
                  Matteo Paltenghi and
                  Jia Le Tian and
                  Michael Pradel and
                  Lingming Zhang},
  title        = {Fuzz4All: Universal Fuzzing with Large Language Models},
  booktitle    = {Proceedings of the 46th {IEEE/ACM} International Conference on Software
                  Engineering, {ICSE} 2024, Lisbon, Portugal, April 14-20, 2024},
  pages        = {126:1--126:13},
  publisher    = {{ACM}},
  year         = {2024},
  url          = {https://doi.org/10.1145/3597503.3639121},
  doi          = {10.1145/3597503.3639121},
  bibsource    = {dblp computer science bibliography, https://dblp.org}
}

@article{non-textual,
  title={Low-Cost and Comprehensive Non-textual Input Fuzzing with LLM-Synthesized Input Generators},
  author={Zhang, Kunpeng and Li, Zongjie and Wu, Daoyuan and Wang, Shuai and Xia, Xin},
  journal={arXiv preprint arXiv:2501.19282},
  year={2025},
  url          = {https://www.usenix.org/system/files/conference/usenixsecurity25/sec25cycle1-prepub-1291-zhang-kunpeng.pdf},
}

@inproceedings{fuzzing-js,
  author       = {Jueon Eom and
                  Seyeon Jeong and
                  Taekyoung Kwon},
  editor       = {Maria Christakis and
                  Michael Pradel},
  title        = {Fuzzing JavaScript Interpreters with Coverage-Guided Reinforcement
                  Learning for LLM-Based Mutation},
  booktitle    = {Proceedings of the 33rd {ACM} {SIGSOFT} International Symposium on
                  Software Testing and Analysis, {ISSTA} 2024, Vienna, Austria, September
                  16-20, 2024},
  pages        = {1656--1668},
  publisher    = {{ACM}},
  year         = {2024},
  url          = {https://doi.org/10.1145/3650212.3680389},
  doi          = {10.1145/3650212.3680389},
  bibsource    = {dblp computer science bibliography, https://dblp.org}
}

@inproceedings{llm-solver,
  title={Hybrid Language Processor Fuzzing via $\{$LLM-Based$\}$ Constraint Solving},
  author={Yang, Yupeng and Yao, Shenglong and Chen, Jizhou and Lee, Wenke},
  booktitle={34th USENIX Security Symposium (USENIX Security 25)},
  pages={6299--6318},
  year={2025},
  url = {https://www.usenix.org/system/files/usenixsecurity25-yang-yupeng.pdf}
}

@inproceedings{Magneto,
  author       = {Zhuotong Zhou and
                  Yongzhuo Yang and
                  Susheng Wu and
                  Yiheng Huang and
                  Bihuan Chen and
                  Xin Peng},
  editor       = {Vladimir Filkov and
                  Baishakhi Ray and
                  Minghui Zhou},
  title        = {Magneto: {A} Step-Wise Approach to Exploit Vulnerabilities in Dependent
                  Libraries via LLM-Empowered Directed Fuzzing},
  booktitle    = {Proceedings of the 39th {IEEE/ACM} International Conference on Automated
                  Software Engineering, {ASE} 2024, Sacramento, CA, USA, October 27
                  - November 1, 2024},
  pages        = {1633--1644},
  publisher    = {{ACM}},
  year         = {2024},
  url          = {https://doi.org/10.1145/3691620.3695531},
  doi          = {10.1145/3691620.3695531},
  bibsource    = {dblp computer science bibliography, https://dblp.org}
}

@inproceedings{PromptFuzz,
  author       = {Yunlong Lyu and
                  Yuxuan Xie and
                  Peng Chen and
                  Hao Chen},
  editor       = {Bo Luo and
                  Xiaojing Liao and
                  Jun Xu and
                  Engin Kirda and
                  David Lie},
  title        = {Prompt Fuzzing for Fuzz Driver Generation},
  booktitle    = {Proceedings of the 2024 on {ACM} {SIGSAC} Conference on Computer and
                  Communications Security, {CCS} 2024, Salt Lake City, UT, USA, October
                  14-18, 2024},
  pages        = {3793--3807},
  publisher    = {{ACM}},
  year         = {2024},
  url          = {https://doi.org/10.1145/3658644.3670396},
  doi          = {10.1145/3658644.3670396},
  bibsource    = {dblp computer science bibliography, https://dblp.org}
}

@inproceedings{TitanFuzz,
  author       = {Yinlin Deng and
                  Chunqiu Steven Xia and
                  Haoran Peng and
                  Chenyuan Yang and
                  Lingming Zhang},
  editor       = {Ren{\'{e}} Just and
                  Gordon Fraser},
  title        = {Large Language Models Are Zero-Shot Fuzzers: Fuzzing Deep-Learning
                  Libraries via Large Language Models},
  booktitle    = {Proceedings of the 32nd {ACM} {SIGSOFT} International Symposium on
                  Software Testing and Analysis, {ISSTA} 2023, Seattle, WA, USA, July
                  17-21, 2023},
  pages        = {423--435},
  publisher    = {{ACM}},
  year         = {2023},
  url          = {https://doi.org/10.1145/3597926.3598067},
  doi          = {10.1145/3597926.3598067},
  bibsource    = {dblp computer science bibliography, https://dblp.org}
}

@inproceedings{ChatAFL,
  author       = {Ruijie Meng and
                  Martin Mirchev and
                  Marcel B{\"{o}}hme and
                  Abhik Roychoudhury},
  title        = {Large Language Model guided Protocol Fuzzing},
  booktitle    = {31st Annual Network and Distributed System Security Symposium, {NDSS}
                  2024, San Diego, California, USA, February 26 - March 1, 2024},
  publisher    = {The Internet Society},
  year         = {2024},
  url          = {https://www.ndss-symposium.org/ndss-paper/large-language-model-guided-protocol-fuzzing/},
  bibsource    = {dblp computer science bibliography, https://dblp.org}
}

@inproceedings{Lyso,
  title={From Alarms to Real Bugs: Multi-target Multi-step Directed Greybox Fuzzing for Static Analysis Result Verification},
  author={Bao, Andrew and Zhao, Wenjia and Wang, Yanhao and Cheng, Yueqiang and McCamant, Stephen and Yew, Pen-Chung},
  booktitle={34th USENIX Security Symposium (USENIX Security 25)},
  pages={6977--6997},
  year={2025},
  url={https://www.usenix.org/system/files/usenixsecurity25-bao-andrew.pdf}
}

@inproceedings{AFLFast,
  author       = {Marcel B{\"{o}}hme and
                  Van{-}Thuan Pham and
                  Abhik Roychoudhury},
  editor       = {Edgar R. Weippl and
                  Stefan Katzenbeisser and
                  Christopher Kruegel and
                  Andrew C. Myers and
                  Shai Halevi},
  title        = {Coverage-based Greybox Fuzzing as Markov Chain},
  booktitle    = {Proceedings of the 2016 {ACM} {SIGSAC} Conference on Computer and
                  Communications Security, Vienna, Austria, October 24-28, 2016},
  pages        = {1032--1043},
  publisher    = {{ACM}},
  year         = {2016},
  url          = {https://doi.org/10.1145/2976749.2978428},
  doi          = {10.1145/2976749.2978428},
  bibsource    = {dblp computer science bibliography, https://dblp.org}
}

@misc{AD,
      title={Attention Distance: A Novel Metric for Directed Fuzzing with Large Language Models}, 
      author={Wang Bin and Ao Yang and Kedan Li and Aofan Liu and Hui Li and Guibo Luo and Weixiang Huang and Yan Zhuang},
      year={2025},
      eprint={2512.19758},
      archivePrefix={arXiv},
      primaryClass={cs.SE},
      url={https://arxiv.org/abs/2512.19758}, 
}

@Article{ISC4DGF,
title = {ISC4DGF: Enhancing Directed Grey-Box Fuzzing with Initial Seed Corpus Generation Driven by Large Language Models},
journal = {Journal of Computer Science and Technology},
volume = {40},
number = {6},
pages = {1662-1677},
year = {2025},
issn = {1000-9000(Print) /1860-4749(Online)},
doi = {10.1007/s11390-025-4745-0},	
url = {https://jcst.ict.ac.cn/en/article/doi/10.1007/s11390-025-4745-0},
author = {Yi-Jiang Xu and Hong-Rui Jia and Li-Guo Chen and Xin Wang and Zheng-Ran Zeng and Yi-Dong Wang and Qing Gao and Wei Ye and Shi-Kun Zhang and Zhong-Hai Wu}
}

@article{evaluatingDGF,
  author       = {Tae Eun Kim and
                  Jaeseung Choi and
                  Seongjae Im and
                  Kihong Heo and
                  Sang Kil Cha},
  title        = {Evaluating Directed Fuzzers: Are We Heading in the Right Direction?},
  journal      = {Proc. {ACM} Softw. Eng.},
  volume       = {1},
  number       = {{FSE}},
  pages        = {316--337},
  year         = {2024},
  url          = {https://doi.org/10.1145/3643741},
  doi          = {10.1145/3643741},
  bibsource    = {dblp computer science bibliography, https://dblp.org}
}

\end{document}